\documentclass[twocolumn]{aastex631}

\def\kpc{{\rm kpc}}

\newcommand{\kms}{{\rm km~s^{-1}}}

\newcommand{\oiii}{[\textrm{O}~\textsc{iii}]}

\newcommand{\simgt}{\,\rlap{\lower 3.5 pt \hbox{$\mathchar \sim$}} \raise
1pt \hbox {$>$}\,}
\newcommand{\simlt}{\,\rlap{\lower 3.5 pt \hbox{$\mathchar \sim$}} \raise
1pt \hbox {$<$}\,}
\newcommand{\Msun}{M_{\odot}}

\newcommand{\logm}{\log M_*/\Msun}

\def\kpc{{\rm kpc}}

\newcommand{\ha}{${\rm H\alpha}$}
\newcommand{\hb}{${\rm H\beta}$}

\newcommand{\Z}{{\rm [Z/H]}}

\newcommand{\nsrc}{341}
\newcommand{\nsrcz}{109}
\newcommand{\nmex}{51}
\newcommand{\ncom}{44}
\newcommand{\ncomspec}{15}
\newcommand{\nfailed}{24} 

\accepted{Dec 4, 2023}

\submitjournal{ApJ}

\shorttitle{Enhanced Sub-kpc Scale Star-formation in Galaxies at $5<z<14$}
\shortauthors{Morishita et al.}

\graphicspath{{./}{figures/}}

\begin{document}

\title{Enhanced Sub-kpc Scale Star-formation: Results From A JWST Size Analysis of \nsrc\ Galaxies At $5<z<14$}

\correspondingauthor{Takahiro Morishita}
\email{takahiro@ipac.caltech.edu}

\author[0000-0002-8512-1404]{Takahiro Morishita}
\affiliation{IPAC, California Institute of Technology, MC 314-6, 1200 E. California Boulevard, Pasadena, CA 91125, USA}

\author[0000-0001-9935-6047]{Massimo Stiavelli}
\affiliation{Space Telescope Science Institute, 3700 San Martin Drive, Baltimore, MD 21218, USA}

\author[0000-0001-7583-0621]{Ranga-Ram Chary}
\affiliation{IPAC, California Institute of Technology, MC 314-6, 1200 E. California Boulevard, Pasadena, CA 91125, USA}

\author[0000-0001-9391-305X]{Michele Trenti}
\affiliation{School of Physics, University of Melbourne, Parkville 3010, VIC, Australia}
\affiliation{ARC Centre of Excellence for All Sky Astrophysics in 3 Dimensions (ASTRO 3D), Australia}

\author[0000-0003-1383-9414]{Pietro~Bergamini}
\affiliation{Dipartimento di Fisica, Università degli Studi di Milano, Via Celoria 16, I-20133 Milano, Italy}
\affiliation{INAF - OAS, Osservatorio di Astrofisica e Scienza dello Spazio di Bologna, via Gobetti 93/3, I-40129 Bologna, Italy}

\author[0000-0003-1564-3802]{Marco Chiaberge}
\affiliation{Space Telescope Science Institute for the European Space Agency (ESA), ESA Office, 3700 San Martin Drive, Baltimore, MD 21218, USA}
\affiliation{The William H. Miller III Department of Physics and Astronomy, Johns Hopkins University, Baltimore, MD 21218, USA}

\author[0000-0003-4570-3159]{Nicha Leethochawalit}
\affiliation{National Astronomical Research Institute of Thailand (NARIT), Mae Rim, Chiang Mai, 50180, Thailand}

\author[0000-0002-4140-1367]{Guido Roberts-Borsani}
\affiliation{Department of Physics and Astronomy, University of California, Los Angeles, 430 Portola Plaza, Los Angeles, CA 90095, USA}
\affiliation{Department of Astronomy, University of Geneva, Chemin Pegasi 51, 1290 Versoix, Switzerland}

\author[0000-0002-6196-823X]{Xuejian Shen}
\affiliation{TAPIR, California Institute of Technology, Pasadena, CA 91125, USA}
\affiliation{Department of Physics \& Kavli Institute for Astrophysics and Space Research, Massachusetts Institute of Technology, Cambridge, MA 02139, USA}

\author[0000-0002-8460-0390]{Tommaso Treu}
\affiliation{Department of Physics and Astronomy, University of California, Los Angeles, 430 Portola Plaza, Los Angeles, CA 90095, USA}



\begin{abstract}
We present a comprehensive search and analysis of high-redshift galaxies in a suite of nine public JWST extragalactic fields taken in Cycle 1
, covering a total effective search area of $\sim358\,{\rm arcmin^2}$. Through conservative ($8\,\sigma$) photometric selection, we identify \nsrc\,galaxies at $5<z<14$, with \nsrcz\ having spectroscopic redshift measurements from the literature, including recent JWST NIRSpec observations. Our regression analysis reveals that the rest-frame UV size-stellar mass relation follows $R_{\rm eff}\propto M_*^{0.19\pm0.03}$, similar to that of star-forming galaxies at $z\sim3$, but scaled down in size by $\sim0.7$\,dex. We find a much slower rate for the average size evolution over the redshift range, $R_{\rm eff}\propto(1+z)^{-0.4\pm0.2}$, than that derived in the literature. A fraction ($\sim13\,\%$) of our sample are marginally resolved even in the NIRCam imaging ($\simlt100$\,pc), located at $\simgt1.5\,\sigma$ below the derived size-mass slope. These compact sources exhibit a high star formation surface density $\Sigma_{\rm SFR}>10\,M_\sun\,{\rm yr^{-1}\,kpc^{-2}}$, a range in which only $<0.01\,\%$ of the local star-forming galaxy sample is found. For those with available NIRSpec data, no evidence of ongoing supermassive black hole accretion is observed. A potential explanation for the observed high \oiii-to-\hb\ ratios could be high shock velocities, likely originating within intense star-forming regions characterized by high $\Sigma_{\rm SFR}$. Lastly, we find that the rest-frame UV and optical sizes of our sample are comparable. Our results are consistent with these early galaxies building up their structures inside-out and yet to exhibit the strong color gradient seen at lower redshift.
\end{abstract}



 \section{Introduction} \label{sec:intro}

In a hierarchical universe, dark matter starts collapsing at initial density peaks, giving rise to the underlying structure. Baryons then start accreting in the dark matter potential wells and forming stars. 
Depending on the initial conditions of gas and dark matter halos, the appearance of the resulting system may differ dramatically \citep[e.g.,][]{mo98,bullock01}. 

Within this context, the size of galaxies is a fundamental and essential proxy for understanding galaxy formation and evolution. Galaxies occupy a relative narrow portion of the size and stellar mass/luminosity plane.  The distribution of galaxies within this so-called size–stellar mass/luminosity relation, the average size growth rate across cosmic time, and the distribution of other structural parameters, such as the S\'ersic index and axis ratio, are key diagnostics of early galaxy formation.

In the local universe, the large statistics enabled by the Sloan Digital Sky survey revealed a fundamental relation of galaxy structures and sizes with mass \citep[e.g.,][]{kauffmann03}. For example, it has been found that local early-type and late-type galaxies follow different slopes in the size-mass plane \citep[e.g.,][]{shen03,guo09,simard11,cappellari13}. Such differences are believed to arise from a combination of initial conditions, evolutionary paths, and environmental influences.

In contrast, observations of galaxies at higher redshifts have revealed a variety of galaxy morphologies \citep{conselice04,wuyts11,guo12,szomoru12}. These observations indicate active physical processes within and between galaxies, establishing the structural sequences seen in the local universe. Prior to JWST, the Hubble Space Telescope (HST) pushed the frontier of the fundamental galaxy size-mass relation to $z\sim7$ \citep{bruce12,mosleh12,vanderwel12,vanderwel14,morishita14,allen17,yang21}. Beyond that redshift, however, the investigation has been severely limited by spatial resolution ($\sim1$\,kpc), as well as the limited number of infrared filters at $>2\,\mu$m, which is critical to robustly infer galaxy stellar masses. As such, the effort beyond the redshift has been largely limited to small samples of relatively luminous galaxies \citep{oesch10,ono13,holwerda15} or a small volume through strong gravitational lens \citep{kawamata15,kawamata18,yang22}

These observable properties are believed to reflect the initial conditions of the gas and dark matter halos from which galaxies form. However, over time, such fundamental properties are susceptible to contamination through a sequence of stochastic, non-linear physical processes, including mergers. Therefore, a detailed characterization of galaxy size becomes imperative, as they may offer insights into not only the physical mechanisms on act, but also their interplay with interstellar medium \citep[e.g.,][]{marshall22,roper22} and the nature of dark matter \citep[e.g.,][]{shen23}.

Early results from Cycle~1 have already demonstrated the remarkable capabilities of JWST, with its red sensitivity and resolution, revealing early galaxy morphologies down to scales of $\simlt100$\,pc, throughout rest-frame UV to optical wavelengths \citep[e.g.,][]{yang22b,naidu22,finkelstein22,treu23,morishita23,huertascompany23,tacchella23,robertson23}. Given the substantial number of observations completed in Cycle~1, this study aims to undertake a comprehensive analysis of galaxy sizes, offering the first large scale systematic study of the galaxy size-mass relation at $z>5$. To achieve this, we perform an analysis based on consistently reduced data from several publicly available extragalactic fields observed during Cycle-1, encompassing a total effective area of $\sim358$\,arcmin$^2$. This extensive coverage enables us to construct a robust sample comprising \nsrc\ galaxies within the redshift range of $5<z<14$. 

The paper is structured as follows: we present our data reduction in Sec.~\ref{sec:data}, followed by our photometric analyses in Sec.~\ref{sec:ana}. We then characterize the structure of identified galaxies and infer the distributions of their structural parameters in Sec.~\ref{sec:res}. We investigate the inferred physical properties and discuss the origin of early galaxies in comparison with lower-$z$ galaxies in Sec.~\ref{sec:disc}. We summarize our key conclusions in Sec.~\ref{sec:sum}. Where relevant, we adopt the AB magnitude system \citep{oke83,fukugita96}, cosmological parameters of $\Omega_m=0.3$, $\Omega_\Lambda=0.7$, $H_0=70\,\kms\, {\rm Mpc}^{-1}$, and the \citet{chabrier03} initial mass function. Distances are in proper units unless otherwise stated.

\section{Data}\label{sec:data}

We base our analysis on nine public deep fields from JWST Cycle~1. For all fields, except for the GLASS/UNCOVER and JADES-GDS fields where the final mosaic images are made publicly available by the teams, we retrieve the raw-level images from the MAST archive and reduce those with the official JWST pipeline, with several customized steps as detailed below. We then apply our photometric pipeline, {\tt borgpipe} \citep{morishita21}, on all mosaic images to consistently extract sources and measure fluxes. Our final high-$z$ source candidates are selected by applying the Lyman-break dropout technique \citep{steidel98} supplemented by photometric-redshift selection, as implemented by \citet{morishita22}.

\subsection{Uniform data reduction of NIRCam images}\label{sec:data-jwst}
In each field, we start with raw ({\tt uncal.fits}) images retrieved from MAST. We use {\tt Grizli} (ver\,1.8.3) to reduce raw images to generate calibrated images ({\tt cal.fits}). In this step, in addition to the official pipeline's {\tt DETECTOR} step, {\tt Grizli} includes additional processes for flagging artifacts, such as snowball halos, claws, and wisps. 

We then apply {\tt bbpn}\footnote{https://github.com/mtakahiro/bbpn} on the calibrated images, to subtract $1/f$-noise. The tool follows the procedure proposed by \citet{schlawin21}. Briefly, {\tt bbpn} first creates object masks; then it calculates background level in each of the four detector segments (each corresponds to the detector amplifiers) and subtracts the estimated background; it then runs through the detector in the vertical direction and again subtracts the background estimated in each row (that consists of 2048\,pixels minus masked pixels); lastly, to compensate for any local over-subtraction of sky near e.g., bright stars or large foreground galaxies, {\tt bbpn} estimates spatially varying background and subtract from the entire image. 

After the $1/f$-noise subtraction step, we align the calibrated images using the {\tt tweakreg} function of the JWST pipeline. For large mosaic fields (i.e. PRIMER and CEERS), we divide the images into subgroups beforehand and process each of those separately, to optimize computing speed and memory usage. In those fields, images are split into subgroups based on the distance of each image to the other images. We here set the maximum distance of $6\,\arcmin$ for images to be in the same subgroup. We ensure that the images taken in the same visits (i.e. eight detector images for the blue channel and two detector images for the red channel of NIRCam) are grouped together, as their relative distance  should remain consistent in the following alignment step.

Images in each subgroup are aligned on a filter-by-filter basis. We provide {\tt tweakreg} a set of images associated with source catalogs, generated by running {\tt SExtractor} \citep{bertin96} on each image. This enables us to eliminate potential artifacts (such as stellar spikes and saturated stars), which are often included by the automated algorithm by {\tt tweakreg}, and secure alignment calculation by using only reliable sources. For all subgroups, each image is aligned to, when available, the WCS reference of a single, contiguous source catalog taken from a large FoV ground-based imaging (see below). This is to avoid alignment issues in some fields and/or subregions, e.g., misalignment in overlapping regions caused by insufficient reference stars. 
It is noted that {\tt tweakreg} estimates a single alignment solution for images that are taken in the same visit and applies it coherently to those images, such that the distance between the imaging detectors remains the same for all visits. 

Once the images are aligned to the global WCS reference, we drizzle and combine the images into a (sub)mosaic using the pipeline step {\tt IMAGE3}. The pixel scale and pixel frac for drizzling are set to 0.0315 and 0.7 for all filters. For the fields that have multiple subgroups, we then create a single mosaic using a python function {\tt reproject}. Lastly, to eliminate any residual shifts, we once again apply {\tt tweakreg} to a set of multi-band mosaics, but based on the source catalog generated by using the F444W mosaic. The images are resampled after the final alignment in the same pixel grid using {\tt reproject}.

\subsection{JWST NIRCam Extragalactic Fields}\label{sec:fields}
To ensure our selection of high-redshift sources is as consistent as possible, we consider fields that have images in at least six filters (F115W, F150W, F200W, F277W, F356W, F444W). Some fields have additional blue filters (F070W, F090W) and several medium bands (F300M, F335M, F410M, F430M, F480M), that extend the search range toward lower redshift and improve photometric redshift estimates. When spectroscopic redshift measurements are available (from either ground or recent JWST observations), we include and use them for photometric flux calibration (Sec.~\ref{sec:photom}), as well as for sample selection (spec-z supercede dropout or photo-z).

\subsubsection{PAR1199}\label{sec:m1149}

The PAR1199 field (11:49:47.31, +22:29:32.1) was taken as part of a Cycle1 GTO program (PID 1199, \citealt{stiavelli23}) in May and June 2023, attached as coordinated parallel to the NIRSpec primary, and released immediately without a proprietary period. The NIRCam imaging consists of eight filters, including F090W and F410M, with a total science exposure of $\sim16$\,hrs with the {\tt MEDIUM8} readout mode. Due toscheduling constraints, the parallel field falls in a field where only shallow (1--2\,orbits) HST ACS F606W and F814W images are available. In addition, because of a bright star that is located near the edge of Module A, the sensitivity limit of the blue channel is slightly shallower in this module compared to the other Module due to the increased scatter light and artifacts. The impact is found to be less in red filters. Galactic reddening is ${E(B-V)} = 0.024$\,mag \citep{schlegel98}. The images are aligned to bright sources in the Pan-STARRS DR2 catalog \citep{chambers16}. The effective area in the detection image is $10.1$\,arcmin$^2$.

\subsubsection{J1235}\label{sec:j1235}
The J1235 field (12:35:54.4631, +04:56:8.50) is one of the low-ecliptic latitude fields that were observed as part of a commissioning program used for NIRCam flatfield (PID 1063, PI Sunnquist). Among the fields available from this program, the J1235 field offers deep NIRCam coverage by 10 filters, including F070W, F090W, F300M, and F480M. Two separate visits were made in March-April 2022 and May 2022. During the first visit, the telescope mirror alignment was not complete, and thus we exclude the data taken during that visit. With the second visit, the exposure time goes as deep as 5.8\,hrs in a single filter (with 50.9\,hrs in total for the entire field), making it one of the deepest NIRCam multi-band fields. Galactic reddening is ${E(B-V)} = 0.024$\,mag.

The total field coverage extends to $\sim34\,{\rm arcmin}^2$ in the effective detection area, about $\sim3.7$ NIRCam footprints; however, some filter images are shifted from others, making the effective area for dropout selection dependent on target redshift. The NIRCam images were aligned to bright sources in the Pan-STARRS DR2 catalog.

\subsubsection{North Ecliptic Pole Time-Domain Field}\label{sec:nep}

The NIRCam imaging in the North Ecliptic Pole Time-domain field \citep[17:22:47.896, +65:49:21.54;][]{jansen18} was taken as part of a GTO program \citep[PID 2738,][]{windhorst23}. The imaging data used here was taken and immediately released after the first epoch of the visit, consisting of eight filters, including F090W and F410M. For the WCS alignment of our JWST data, we use a publicly available catalog that consists of sources observed with the Subaru/HSC instrument \citep{miyazaki18} as part of the Hawaii EROsita Ecliptic pole Survey (HEROES, PI G. Hasinger \& E. Hu).\footnote{\url{http://lambda.la.asu.edu/jwst/neptdf/Subaru/index.html}} Galactic reddening is ${E(B-V)} \sim 0.028$\,mag. The effective area in the detection image is $16.9$\,arcmin$^2$.

\subsubsection{Primer-UDS and COSMOS}\label{sec:primer}
Two large NIRCam mosaics are scheduled in a Cycle 1 GO program, Public Release Imaging For Extragalactic Research (PRIMER, PID 1837, PI Dunlop). PRIMER observed two extragalactic fields, the CANDELS UDS and COSMOS fields \citep{grogin11,koekemoer11}. The visits are configured with a consistent set of filters (eight filters, including F090W and F410M) and exposure time. In this study, we use the data taken during the first visit in both fields, which covers most of the entire planned fields. However, several images were identified as failed guide star acquisitions and were removed from our reduction.

The UDS mosaics are aligned to the SPLAXH SXDS catalog \citep{mehta18}. We include spec-$z$ measurements available in the same catalog. The COSMOS mosaics is aligned to the COSMOS2020 catalog \citep{weaver22}. We include spec-$z$ measurements made by the Keck/DEIMOS instrument and published in \citet{hasinger18}. Galactic reddening is ${E(B-V)} \sim 0.023$\,mag and $0.017$\,mag, respectively. The effective area in the detection images are $128.2$\,arcmin$^2$ and $93.9$\,arcmin$^2$.

\subsubsection{New Generation Deep Field}\label{sec:ngdeep}
The Next Generation Deep Extragalactic Exploratory Public survey, NGDEEP \citep[PID 2079, PI Finkelstein;][]{bagley23}, is a deep spectroscopic+imaging program using the NIRISS WFSS as the primary mode and the NIRCam imaging attached as coordinated parallel in the HUDF-Par2 field \citep{stiavelli05hst,oesch07,illingworth09hst}. In this study, we use the epoch1 NIRCam imaging data, whereas the epoch2 imaging is currently scheduled in early 2024. 

The NIRCam field consists of six filters. Due to the use of the {\tt DEEP8} readout mode for many of the NIRCam exposures, a small portion of the final images are severely contaminated by CR hits, which moderately reduces the effective field area and increases the contamination fraction in the high-$z$ source selection. We include spec-$z$ measurements made by the VANDELS collaboration \citep{pentericci18}. Galactic reddening is ${E(B-V)} \sim 0.007$\,mag. The effective area in the detection image is $10.2$\,arcmin$^2$.

\subsubsection{CEERS}\label{sec:ceers}
The Cosmic Evolution Early Release Science Survey, or CEERS \citep{bagley23ceers,finkelstein23}, is an ERS program (PID 1345, PI Finkelstein). The data set consists of eight NIRCam filters in the EGS field, previously studied with HST, including as part of CANDELS. The images are aligned to the WCS of the HST F606W image released by the CEERS team ({\tt HDR1}), which is originally aligned to the GAIA-EDR3 WCS.

The CEERS observations had two separate visits, one in June 2022 (for four sub-regions, \#1, 2, 3, 6) and the other in December 2022 (\#4, 5, 7, 8, 9, 10). We reduce the NIRCam images separately in each subfield for the reason described in Sec.~\ref{sec:data-jwst}. Galactic reddening is ${E(B-V)} \sim 0.010$\,mag.

We include spec-$z$ measurements from multiple studies \citep{skelton14,momcheva16,roberts-borsani16,larson22}, as well as recent JWST spectroscopy studies \citep{arrabalhalo23,arrabalharo23b,harikane23,harikane23c,fujimoto23,larson23,kocevski23,nakajima23,sanders23,tang23}. The effective area in the detection image is $117.5$\,arcmin$^2$.

\subsubsection{GLASS-JWST/UNCOVER}\label{sec:a2744}
Multiple Cycle 1 programs observed the Abell 2744 field (00:14:21, -30:24:03), including the GLASS-JWST Early Release Science Program \citep[PID~1324;][]{treu22}, a Treasury Survey program the Ultradeep NIRSpec and NIRCam ObserVations before the Epoch of Reionization \citep[UNCOVER, PID 2561;][]{bezanson22}, and a JWST DDT program \citep[PID~2756; PI. W. Chen;][]{roberts-borsani22b}. 
In this study, we utilize the public imaging data made available by the GLASS-JWST team \citep{merlin22,paris23}. Their reduction processes include several customized steps, to eliminate detector artifacts. The data set consists of eight NIRCam filters, including F090W and F410M. We use the public lens model by \citet{bergamini23b} to correct lens magnification of the background sources. We include spectroscopic measurements made available in the literature \citep[][]{braglia09,owers11,schmidt14,richard21,bergamini23}, including those from recent JWST observations \citep{roberts-borsani22,roberts-borsani22b,morishita23b,mascia23,jones23}. Galactic reddening is $E(B-V)\sim0.013$\,mag. The effective area in the detection image is $48.4$\,arcmin$^2$. We hereafter refer to the field as A2744 for the sake of simplicity.

\subsubsection{JADES-GDS}\label{sec:jades-s}
We include a deep field from the JWST Advanced Deep Extragalactic Survey \citep[JADES;][]{robertson23,tacchella23,eisenstein23}. As of the time of writing, NIRCam imaging data in one of the deep fields in the GOODS-South field (3:32:39.3, -27:46:59) are publicly available \citep{hainline23,rieke23}. We retrieve the fully processed images and spectroscopic catalogs made available by the team. The data set consists of nine NIRCam filters, including F090W, F335M, and F410M. We include spectroscopic sources listed in the MSA spectroscopic catalog \citep[][]{bunker23},
as well as those from the VANDELS survey \citep{pentericci18}. Galactic reddening is ${E(B-V)} \sim 0.007$\,mag. The effective area in the detection image is $26.7$\,arcmin$^2$.

\begin{deluxetable*}{ccccccccccccc}
\tablecaption{
$5\,\sigma$-limiting magnitudes (in ABmag).
}
\tablehead{\colhead{Field ID} & \colhead{F070W} & \colhead{F090W} & \colhead{F115W} & \colhead{F150W} & \colhead{F200W} & \colhead{F277W} & \colhead{F300M} & \colhead{F335M} & \colhead{F356W} & \colhead{F410M} & \colhead{F444W} & \colhead{F480M}
}
\startdata
PAR1199 & -- & 29.0 & 29.0 & 29.0 & 29.2 & 29.4 & -- & -- & 29.5 & 28.9 & 28.9 & -- \\
J1235 & 28.6 & 28.6 & 28.6 & 28.3 & 28.9 & 29.1 & 28.8 & -- & 29.2 & -- & 28.3 & 27.7 \\
NEP & -- & 28.4 & 28.5 & 28.6 & 28.7 & 29.0 & -- & -- & 29.2 & 28.4 & 28.6 & -- \\
PRIMERUDS$^\dagger$ & -- & 27.8 & 27.8 & 28.0 & 28.1 & 28.4 & -- & -- & 28.4 & 27.6 & 27.9 & -- \\
PRIMERCOS$^\dagger$ & -- & 27.8 & 27.8 & 28.0 & 28.1 & 28.4 & -- & -- & 28.5 & 27.8 & 28.2 & -- \\
NGDEEP & -- & -- & 29.8 & 29.6 & 29.7 & 29.6 & -- & -- & 29.7 & -- & 29.2 & -- \\
CEERS & -- & -- & 28.7 & 28.6 & 28.8 & 29.0 & -- & -- & 29.0 & 28.2 & 28.5 & -- \\
A2744 & -- & 29.2 & 28.9 & 28.8 & 28.8 & 29.0 & -- & -- & 29.1 & 28.5 & 28.9 & -- \\
JADESGDS & -- & 29.6 & 29.9 & 29.8 & 29.9 & 30.2 & -- & 29.6 & 30.1 & 29.6 & 29.8 & --
\enddata
\tablecomments{
Limiting magnitudes measured in empty regions of the image with $r=0.\!''16$ apertures. $^\dagger$: Mosaic images have been created using the first epoch data that are available as of June 2023.
}\label{tab:limmag}
\end{deluxetable*}

\begin{deluxetable}{ccccc}
    \tablecaption{
    Numbers of the final sources in fields.
    }
    \tablehead{
    \colhead{Field} & \colhead{F070W-d} & \colhead{F090W-d} & \colhead{F115W-d} & \colhead{F150W-d}
    }
    \startdata
PAR1199 & 0 (0) & 7 (0) & 0 (0) & 0 (0)\\
J1235 & 20 (0) & 1 (0) & 0 (0) & 0 (0)\\
NEP & 0 (0) & 11 (0) & 0 (0) & 0 (0)\\
PRIMERUDS & 0 (0) & 44 (0) & 10 (0) & 0 (0)\\
PRIMERCOS & 2 (2) & 24 (0) & 3 (0) & 1 (0)\\
NGDEEP & 1 (1) & 0 (0) & 1 (0) & 0 (0)\\
CEERS & 33 (33) & 8 (8) & 9 (2) & 0 (0)\\
A2744 & 19 (19) & 15 (9) & 10 (1) & 0 (0)\\
JADESGDS & 26 (26) & 86 (5) & 9 (2) & 1 (1)\\
\hline
ALL & 101 (81) & 196 (22) & 42 (5) & 2 (1)\\
    \enddata
    \tablecomments{
    ``-d'' represents dropout. Numbers of spectroscopically confirmed sources ({see Sec.~\ref{sec:spec}}) are shown in brackets. 
    }\label{tab:nsrc}
    \end{deluxetable}

\section{Analysis}\label{sec:ana}

\subsection{Photometry} 
\label{sec:photom}

The photometric catalog in each field is constructed following \citet{morishita22}, using {\tt borgpipe} \citep{morishita21}. Briefly, we first prepare a detection image for each field by stacking the F277W, F356W, and F444W filters weighted by each of their RMS maps. Source identification is made in the detection image using SExtractor \citep{bertin96}. Fluxes are then estimated for the detected sources with a $r=0.\!''16$ aperture. For the aperture flux measurement, images are PSF-matched to the PSF size of F444W beforehand. The image convolution kernels are generated by using {\tt pypher} \citep{boucaud16} on the psf images generated by {\tt webbpsf} \citep{perrin14}. 

We follow the standard procedure used in the literature \citep{trenti12,bradley12,calvi16,morishita18b,morishita22}, including correction for Galactic extinction and RMS scaling that accounts for correlation noise in drizzled images. Limiting magnitudes of the images are measured by using the same aperture size and reported in Table~\ref{tab:limmag}. 
Aperture fluxes of individual sources are then corrected by applying the correction factor $C = f_{\rm auto, F444W}/f_{\rm aper, F444W}$ universally to all filters, where $f_{\rm auto, F444W}$ is {\rm FLUX\_AUTO} of SExtractor, measured for individual sources. With this approach, colors remain those measured in apertures, whereas the total measurements derived in the following analyses (such as stellar mass, and star formation rate) represent the total fluxes (see also Sec~\ref{sec:size}).

Lastly, since several fields have a number of spectroscopic objects across a wide redshift range, we run {\tt eazypy}, a python wrapper of photometric redshift code {\tt eazy} \citep{brammer08}, to fine-tune fluxes across all filters. A set of correction factors for all filters is derived in each field, from the redshift fitting results on those with spectroscopic redshift. The derived correction factors are found to be $<2\,\%$ relative to the pivot filter, here set to F150W in all fields, only requiring minor correction.

\subsection{Selection of high-redshift galaxy candidates}\label{sec:sel}
In what follows, we present our selection of high-redshift galaxies and galaxy candidates. To construct a robust photometric sample, we adopt a two-fold selection method, which has been established in our previous studies \citep{morishita18b,roberts-borsani22borg,ishikawa22}, and is described in the following two subsections. 

\subsubsection{Lyman-break dropout selection}\label{sec:drop}
Here we identify dropout sources in four redshift ranges. For those detected at Signal-to-noise ratio (S/N) $>4$ in the detection band, we apply one of the following criteria:

{\bf F070W-dropouts ($5.0<z<7.2$)}
$$S/N_{\rm 115} > 8 $$
$$S/N_{\rm 070} < 2.0$$
$$z_{\rm set} = 4.5$$

{\bf F090W-dropouts ($7.2<z<9.7$)}
$$S/N_{\rm 150} > 8 $$
$$S/N_{\rm 090, 070} < 2.0$$
$$z_{\rm set} = 6$$

{\bf F115W-dropouts ($9.7<z<13.0$)}
$$S/N_{\rm 200} > 8 $$
$$S/N_{\rm 115, 090, 070} < 2.0$$
$$z_{\rm set} = 8$$

{\bf F150W-dropouts ($13.0<z<18.6$)}
$$S/N_{\rm 277} > 8 $$
$$S/N_{\rm 150, 115, 090, 070} < 2.0$$
$$z_{\rm set} = 10$$
where S/N is measured in a $r=0.\!''16$ aperture. In each redshift range, we ensure secure selection by requiring a ${\rm S/N}>8$ detection in a filter that covers rest-frame UV ($\sim1600\,\mathrm{\AA}$, but not including the blue side of Lyman break). 
{This stringent requirement ensures high completeness ($>50\,\%$) and reliable size measurements (Appendix~C).}
For the source to be selected as dropout, we require non-detection of fluxes (${\rm S/N}<2$) in all available dropout filters (listed as subscripts above). Furthermore, to secure the non-detection, we repeat the non-detection step with a smaller aperture, $r=0.\!''08$ ($\sim2.5$\,pixel). {Note that a photometric selection is not attempted for fields where no dropout filter is available (but see Sec.~\ref{sec:spec}).} 

A major difference from the conventional Lyman break technique in the literature \citep[e.g.,][]{bouwens23} is that our selection method above does not cut samples based on the color of the rest-frame UV but only on the strength of Lyman break. The choice is made to preserve as many potential sources as possible and make the selection comprehensive --- for example, in a conventional color-cut selection, sources may be dismissed for their color that {\it barely} miss the selection window, even if the color is consistent within the photometric uncertainty. This also means that the fraction of low-$z$ interlopers misidentified as high-$z$ sources is likely increased with regard to the standard technique. Therefore, we further secure the sample in the following step.

\subsubsection{Photometric-redshift selection}\label{sec:photz}
We here secure the dropout sources by applying photometric-redshift selection to the dropout sources selected above. This is to minimize the fraction of low-$z$ interlopers, such as galaxies of relatively old stellar populations \citep[e.g.,][]{oesch16} and with dust extinction, or foreground dwarfs {\citep[e.g.,][]{morishita20}}. Such interlopers are often distinguished by their distinctive red color, readily discernible in our wavelength coverage with NIRCam.

To estimate photometric redshifts, we run {\tt eazy} with the default magnitude prior \citep[{Fig.~4 in}][]{brammer08}. The fitting redshift range is set to $0<z<30$, with a step size of $\log (1+z) = 0.01$.  By comparing photometric redshifts with spectroscopic ones, we find that the template library provided by \citet{hainline23} offers an improved photometric redshift accuracy over the default (v1.3) template library, and thus in this work we adopt the former.

Following \citet{morishita18b}, we exclude sources that satisfy $p(z<z_{\rm set})>0.2$, i.e. total redshift probability at $z<z_{\rm set}$ is greater than $20\,\%$, where $z_{\rm set}$ is set separately for each selection redshift range (see Sec.~\ref{sec:drop}). To eliminate potential contamination by cool (T/L/M-type) stars (i.e. brown dwarfs), we follow \citet{morishita21} and repeat the phot-$z$ analysis with dwarf templates. A set of dwarf templates taken from the IRTF spectral Library \citep{rayner03} is provided to {\tt eazy} and fit to the data with redshift fixed to 0. The fitting result is inferred for every photometric source that is unresolved (see Sec.~\ref{sec:compact}), and the source is removed if the $\chi^2/\nu$ value is smaller than the one from the galaxy template fitting above. {We have excluded 60 sources in this step.}

Lastly, we visually inspect all sources that pass the two selections above. In this step, we exclude any suspicious sources whose flux measurements may be significantly affected, including those with residuals of CRs, close to/overlapping with a brighter galaxy (caused by deblending), misidentified stellar spikes, and those near the detector edge where a part of the source is truncated. {We have discarded 342 sources.} 

{
\subsubsection{Spectroscopic Sample}\label{sec:spec}
In addition to the photometric sample constructed above, we include those with spectroscopic redshift confirmed by previous spectroscopic observations as described in Sec.~\ref{sec:fields}. We add sources when their spectroscopic redshift is within the redshift range defined for each selection window {\it and} when they satisfy detection criteria in the detection (S/N\,$>4$) and rest-frame UV ($>8$) bands. 

The addition of spec-$z$ sources aids in particular the F070W-dropout sample, which would need F070W as a non-detection filter. All fields except for J1235 do not have the filter coverage, leaving the sample size relatively small without spectroscopically confirmed objects. On the other hand, this could introduce a potential bias toward strong line emitters. However, in Sec.~\ref{sec:size-mass} we investigate this in our size-mass analysis and find that the addition of spec-$z$ sources does not impact any of our final conclusions. 
}

\subsection{Size measurements}\label{sec:size}

Our primary analysis is based on the size measurement of galaxies. Following standard practice, we adopt the S$\acute{\mathrm{e}}$rsic profile \citep[][]{sersic63}:

\begin{eqnarray}
I(r)\propto \exp[-b_n(\frac{r}{R_{\mathrm{e}}})^{1/n}-1]
\end{eqnarray}

\noindent
where size is characterized by the effective radius, $R_\mathrm{e}$, which encloses half of the total light of the galaxy. $n$ is the S$\acute{\mathrm{e}}$rsic index and $b_n$ is an $n$-dependent normalization parameter. We model the two-dimensional light profile of each galaxy using {\tt galfit} \citep{peng02, peng10}. 

We follow \cite{morishita14,morishita17} for detailed procedures, with a few modifications to accommodate efficiency and accuracy. Briefly, for each galaxy, we first generate image cutouts (here set to $151 \times 151$\,pixel in size, equivalent to $4.\!''8 \times 4.\!''8$) of original (i.e. pre-psf matched) science map, rms map, and segmentation map. We fix S\'ersic index $n$ to 1, a value that is found to offer reasonable fit to high-$z$ Lyman-break galaxies in the literature \citep{shibuya15,yang22,ono22}. As a test, we repeated the analysis with $n$ as a free parameter {and indeed found its distribution to be centered around $\sim1$. However, this led to an increased fraction ($\sim13\,\%$) of unsuccessful fits, where the solutions either did not converge or converged to unrealistic parameters (i.e. $n<0.2$ or $n>10$).} 
To mitigate potential bias from this constraint, we reevaluate the uncertainty of each size measurement by adding the difference in $R_e$ resulting from the two procedures ($n$-fixed and $n$-free) in quadrature. Consequently, those with $n$ deviating significantly from 1 incur a larger uncertainty in the size measurement.

\begin{figure}
\centering
	\includegraphics[width=0.49\textwidth]{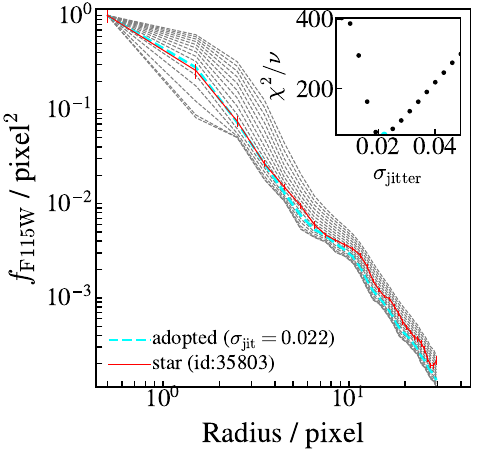}
	\caption{
    Radial flux profiles of {\tt webbpsf} with various $\sigma_{\rm jitter}$ values (dashed lines) generated for the JADESGDS F115W image, where the adopted profile ($\sigma_{\rm jitter}=0.022$) is color-coded in cyan. The profile of an example bright star taken from the image is shown for comparison (red solid line). The inset shows the distribution of $\chi^2/\nu$ of the same star fitted by the {\tt webbpsf} for different $\sigma_{\rm jitter}$ values.
    }
\label{fig:psf}
\end{figure}

The psf image generated by {\tt webbpsf} of the corresponding filter is fed to {\tt galfit}, for convolution of the model profile at each iteration. The psf image is generated for each field by retrieving the Optical Path Difference files of the observed date \footnote{https://webbpsf.readthedocs.io/en/latest/available\_opds.html}. As reported in several studies \citep[][]{ono22,tacchella23,ito23}, we find that the default output of {\tt webbpsf} exhibits a narrower psf profile compared to observed stars in our reduced images. This is potentially caused by e.g., drizzling/resampling of the actual images, as well as jitter in pointing, which could affect more significantly in a longer exposure. We therefore find an optimal psf that describes the actual psf size of our image by tweaking the {\tt jitter\_sigma} parameter of {\tt webbpsf}. To do this, we visually identify unsaturated, bright stars that do not have any companion within $<50$\,pixels. We then fit these stars with various psf models by using {\tt galfit} and select a model that offers the minimum $\chi^2$ value. While it is ideal to repeat the analysis and determine the jitter value in each field, some fields do not have sufficient number of stars that can be used for this. We thus adopt the median value that is determined by stars in all fields for each filter. Figure~\ref{fig:psf} shows an example radial flux profile of an actual star, compared with those of {\tt webbpsf} generated with different $\sigma_{\rm jitter}$ values. The final {\tt jitter\_sigma} value is set $\sim0.022$\,arcsec for the blue-channel filters and $\sim0.034$\,arcsec for the redder-channel.

{Neighboring sources that are close to and relatively bright compared to the main galaxy are fit simultaneously, while the rest of the sources in the stamp are masked using the segmentation map generated by {\tt SExtractor} above. We include any neighbouring source at distance $d$ from the main galaxy when its flux is above the limiting flux defined as:}
\begin{equation}
    f_{\rm nei,lim} = \gamma f_{\rm main} (d/d_{\rm s})^{\alpha_{\rm nei}},
\end{equation}
{with $\gamma=0.8$, $\alpha_{\rm nei}=2.0$, and $d_{\rm s}=60$\,pixel.}

We run {\tt galfit} in the order of the target source magnitudes. The fitting results of the primary galaxy are continuously stored, so that the parameters for the repeated galaxies are fixed to the previously determined values when they appear in the cutout of a fainter galaxy later in the fitting session.

For each galaxy, we repeat the fit in two filters that corresponds to rest-frame UV and optical wavelengths. We then inspect all fitting results to ensure the measurements. We have flagged \nfailed\ sources that show significant residuals, e.g., from multiple clumps within the defined segment region and/or clear features of interaction with nearby sources. These flagged sources are excluded from statistical analyses in the following sections. The measured sizes are presented in Appendix.

\begin{figure}
\centering
	\includegraphics[width=0.5\textwidth]{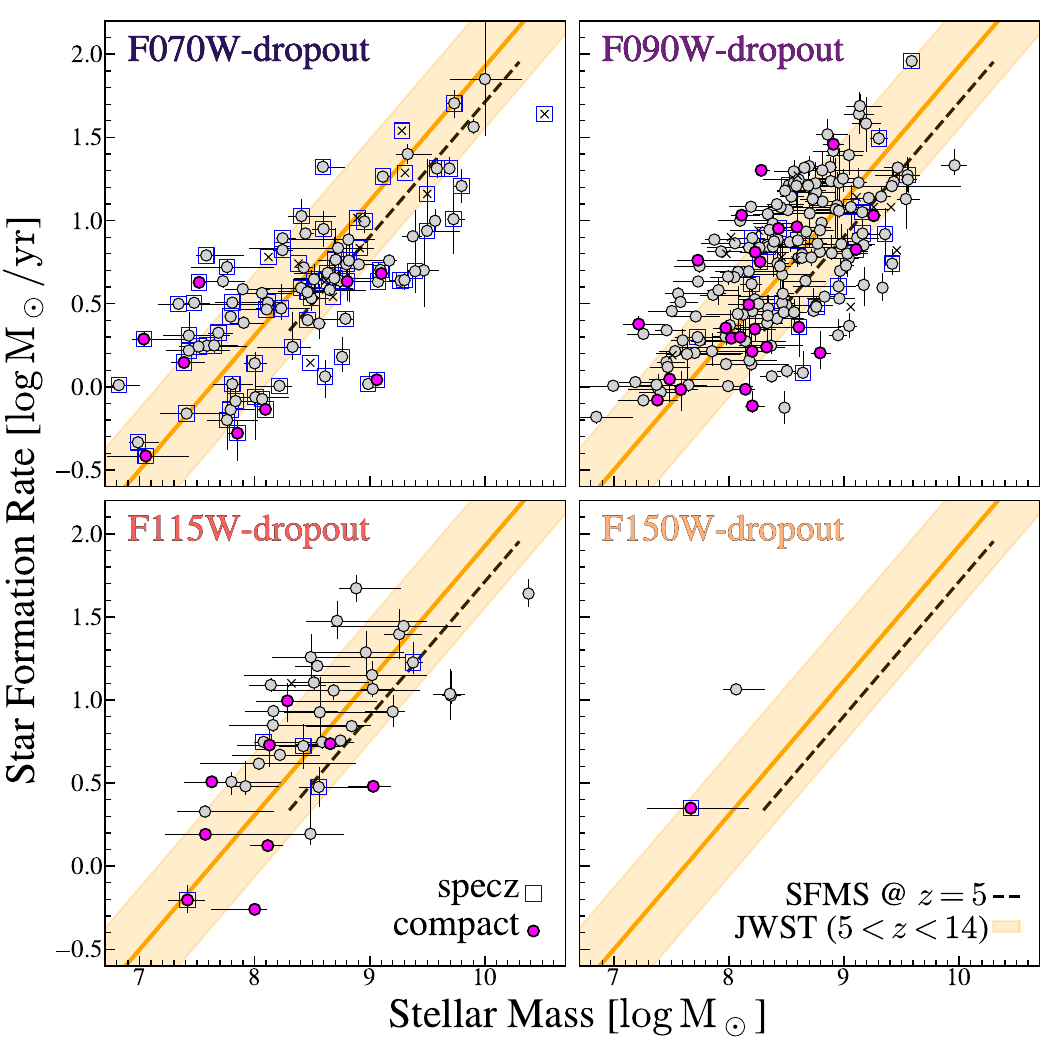}
	\caption{
    Sample distribution in the star formation rate--stellar mass plane (circles). Those with spectroscopic redshift (blue square) and flagged as compact (magenta; Sec~\ref{sec:compact}) are marked accordingly. Those flagged in galfit results (Sec.~\ref{sec:size}) are shown by open symbols. The SFMS slope at $z=5$ (black dashed lines; \citealt{speagle14}), the linear regression slope derived for our entire sample, $\log {\rm SFR} = 0.81 \log (M_*/10^{8}\,M_\odot) + 0.31$ (with the slope fixed to the \citealt{speagle14}'s at $z=5$), and the derived range for the intrinsic scatter ($0.37$\,dex; shaded regions) are shown.
    }
\label{fig:sfms}
\end{figure}

\subsection{Physical Properties inferred by spectral energy distribution analysis}\label{sec:sed}
We infer the spectral energy distribution (SED) of the individual galaxies through SED fitting using photometric data that covers 0.6--5\,$\mu$m. We use the SED fitting code {\tt gsf} \citep[ver1.8;][]{morishita19}, which allows flexible determinations of the SED by adopting binned star formation histories, also known as {\it non-parametric}. {\tt gsf} determines an optimal combination of stellar and interstellar medium (ISM) templates among the template library. For this study, we generate templates of different ages, [10, 30, 100, 300, 1000, 3000]\,Myrs, {and metallicities $\log Z_*/Z_\odot \in [-2, 0]$ at an increment of 0.1} by using {\tt fsps} \citep{conroy09fsps,foreman14}. A nebular component (emission lines and continuum) that is characterized by an ionization parameter {$\log U \in [-3, -1]$} is also generated by {\tt fsps} {\citep[see also][]{byler17}} and added to the template after multiplication by an amplitude parameter. Dust attenuation and metallicity of the stellar templates are treated as free parameters during the fit, whereas the metallicity of the nebular component is synchronized with the metallicity of the stellar component during the fitting process.

The posterior distribution of the parameters is sampled by using \texttt{emcee} for $10^4$ iterations with the number of walkers set to 100. The final posterior is collected after excluding the first half of the realizations (known as burn-in). The resulting physical parameters (such as stellar mass, star formation rate, rest-frame UV slope $\beta_{\rm UV}$, metallicity, dust attenuation $A_V$, and mass-weighted age) are quoted as the median of the posterior distribution, along with uncertainties measured from the 16\,th to 84\,th percentile range. The star-formation rate of individual galaxies is calculated with the rest-frame UV luminosity ($\sim1600\,\rm{\AA}$) using the posterior SED. The UV luminosity is corrected for dust attenuation using the $\beta_{\rm UV}$ slope, which is measured by using the posterior SED, as in \citet{smit16}:
\begin{equation}
    A_{1600} = 4.43 + 1.99\,\beta_{\rm UV}
\end{equation}
The attenuation corrected UV luminosity is then converted to SFR via the relation in  \citet{kennicutt98}:
\begin{equation}
    {\rm SFR\,[M_\odot\,yr^{-1}]} = 1.4 \times 10^{-28} L_{\rm UV}\,[{\rm erg\,s^{-1}\,Hz^{-1}}].
\end{equation}

Lastly, we correct both stellar mass and star formation rate measurements to the total model magnitude derived by {\tt galfit}, as in \citet{morishita14}, by multiplying the correction factor:
\begin{equation}
    C_{\rm galfit} = 10^{-0.4 (m_{\rm galfit}-m_{\rm total})}
\end{equation}
where $m_{\rm galfit}$ is the best-fit total magnitude derived by {\tt galfit} and $m_{\rm total}$ is the total magnitude derived in Sec.~\ref{sec:photom}, both measured in the rest-frame UV filter of interest for the target redshift range. Sources flagged in the {\tt galfit} results are set $C_{\rm galfit} = 1$. The inferred physical properties are presented in Appendix.

\begin{figure}
\centering
    \includegraphics[width=0.49\textwidth]{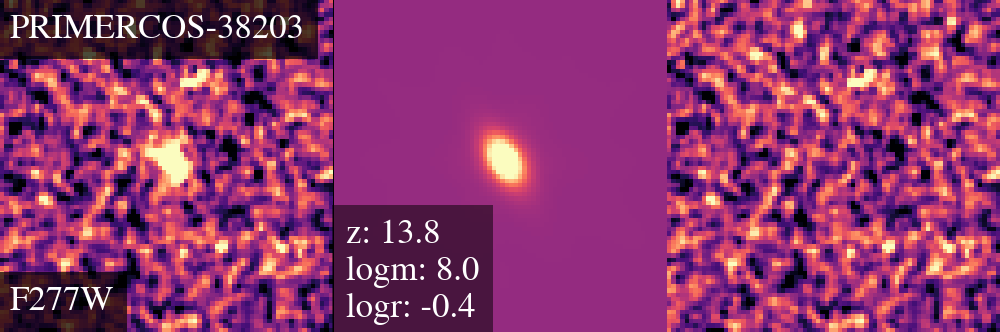}
    \includegraphics[width=0.49\textwidth]{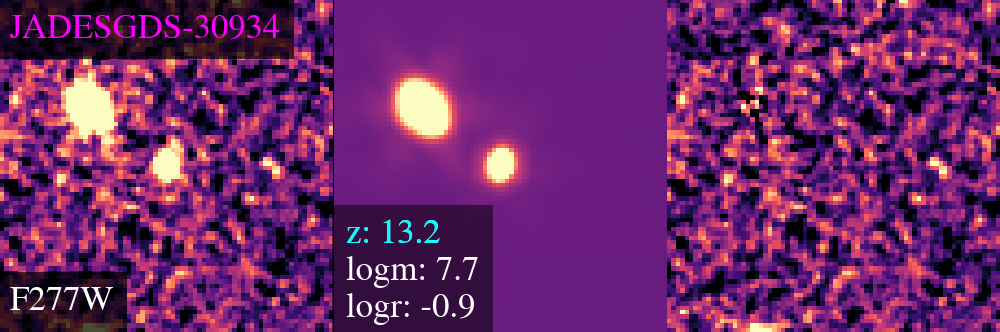}
    \includegraphics[width=0.49\textwidth]{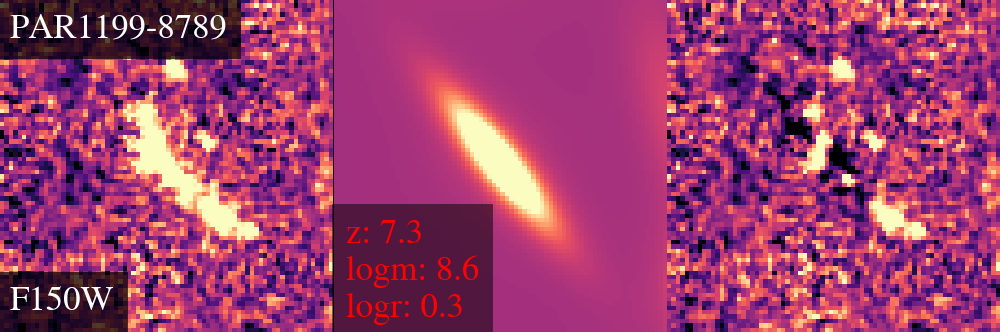}
	\caption{
 Example of {\tt galfit} fitting results in 2D-images (left: original, middle: model, right: residual). (top row): PRIMERCOS-38203, one of the highest redshift sources in our sample. (middle row): JADESGDS-30934 (spectroscopically confirmed to $z=13.2$; \citealt{curtis-lake23}), an example of those classified as compact (Sec.~\ref{sec:compact}). (bottom row): F090W-dropout source that is flagged in {\tt galfit} fitting analysis. Flagged objects are not included in our statistical discussion.
	}
\label{fig:galfit}
\end{figure}

\begin{figure*}
\centering
	\includegraphics[width=0.88\textwidth]{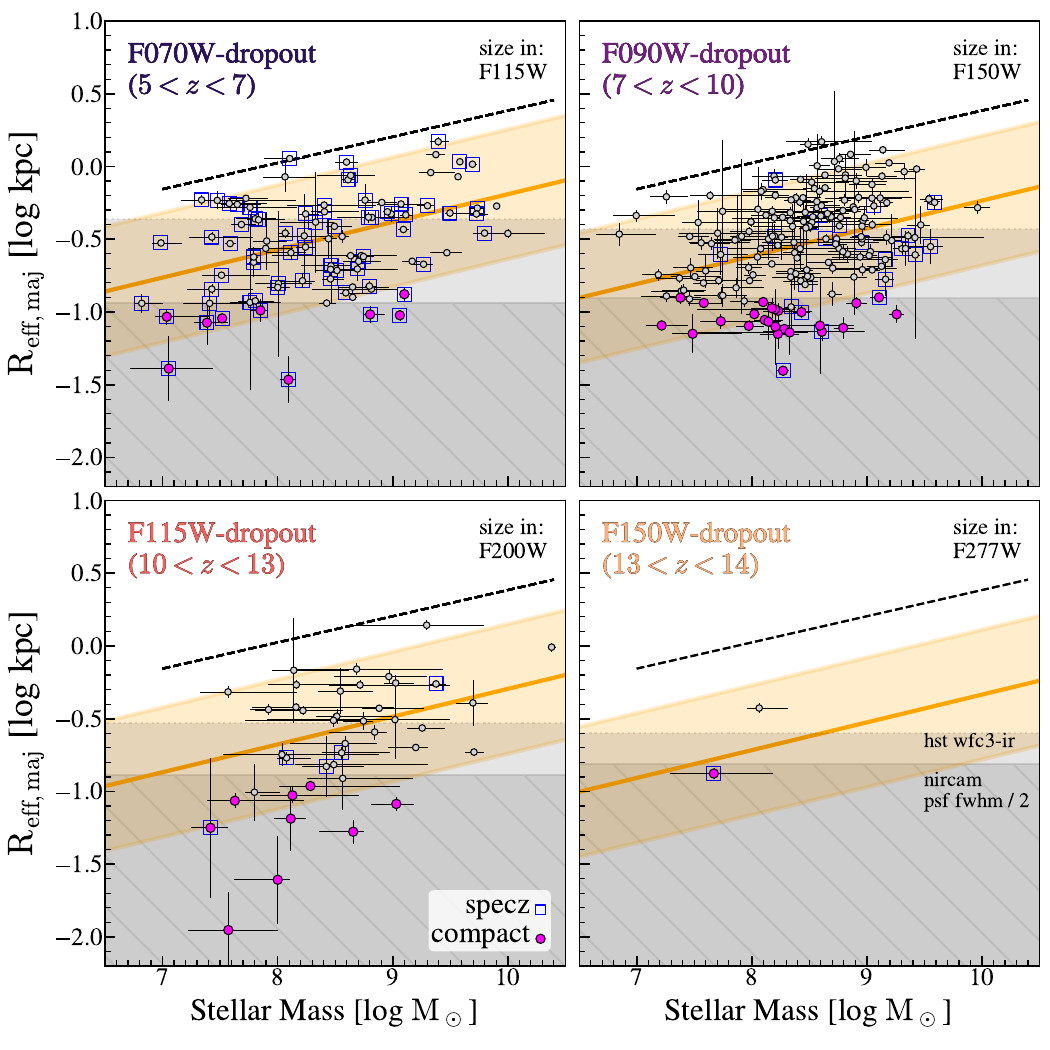}
	\caption{
    Distribution of our sample galaxies at $5<z<14$ in the stellar mass-size plane (gray circles), in four redshift panels. The linear slope derived by regression analysis for the full sample (orange solid lines, with the shaded regions covering the $\pm1.5\,\sigma_{\log R_e}$ range from the median slope; Table~\ref{tab:param}) is shown. For comparison, the slope derived for late-type galaxies at $z\sim2.7$ \citep[][black dashed lines]{vanderwel14} is shown. Those classified as compact (Sec.~\ref{sec:compact}) are shown by magenta symbols. Those with spectroscopic redshift are marked by open blue squares. Two horizontal hatched regions show the physical size of FWHM/2 for NIRCam (dark gray) and HST/WFC3-IR F160W (light gray) at the median redshift of the sources in each redshift window. 
	}
\label{fig:sizemass}
\end{figure*}

\tabletypesize{\footnotesize} \tabcolsep=0.11cm
\begin{deluxetable*}{@{\extracolsep{4pt}}rrrrrrrrrr@{}} 
    \tablecolumns{10}
    \tablewidth{0pt}
    \tablecaption{Size--Mass Relations of Galaxies at $5<z<14$: Linear Regression Best-Fit Coefficients}
    \tablehead{
    \multicolumn{5}{c}{All sample} & \multicolumn{5}{c}{Resolved sample ($R_e>{\rm FWHM/2}$)} \\
    \cline{1-5}
    \cline{6-10}
    \colhead{$N$} & \colhead{$\alpha$} &
    \colhead{$\beta_z$} & \colhead{$\alpha_z$} & \colhead{$\sigma_\mathrm{\log R_e}$} &  \colhead{$N$} & \colhead{$\alpha$} & \colhead{$\beta_z$} & \colhead{$\alpha_z$} & \colhead{$\sigma_\mathrm{\log R_e}$}
    }
\startdata
  317  & $0.19_{-0.03}^{+0.03}$ & $-0.21_{-0.20}^{+0.20}$ & $-0.44_{-0.21}^{+0.21}$ & $0.30_{-0.01}^{+0.01}$ & 278  & [$0.19$] & $-0.33_{-0.18}^{+0.18}$ & $-0.24_{-0.20}^{+0.20}$ & $0.25_{-0.01}^{+0.01}$
  \enddata   
\tablecomments{
Best-fit coefficients for the single slope regression, $\log R_{\rm e}/\kpc=\alpha \log (M_*/10^{8} \Msun) + B(z)$, where $B(z) = \beta_z + \alpha_z \log(1+z)$. Note that $\sigma_{\log R_e} = \sigma_{\ln R_e} / \ln(10)$.
}
\label{tab:param}
\end{deluxetable*}

\section{Results}\label{sec:res}

\subsection{Overview of the final sample}\label{sec:sample}
In Table~\ref{tab:nsrc}, we report the number of our final sources selected in each field and dropout selection window. From all fields, we identify 101 F070W-dropout sources (81 of which are spec-$z$ confirmed), 196 F090W-dropout sources (22), 42 F115W-dropout sources (5), and 2 F150W-dropout sources (1). As we present in the following sections, the sample spans a wide range of stellar mass ($6.8 \simlt \logm \simlt 10.4$) and absolute UV magnitude ($-23 \simlt M_{\rm UV}\,/\,{\rm mag} \simlt -17$). {The final sources are identical when a larger aperture ($0.\!''32$) is adopted for the source selection in Sec.~\ref{sec:drop}.}

In Fig.~\ref{fig:sfms}, we show the distribution of the final sample in the stellar mass-star formation rate plane. Despite the wide mass (more three orders of magnitude) and redshift ($5<z<14$) ranges, our galaxies are found to distribute along a sequence, suggesting that most of our sample galaxies are a typical star-forming population. To compare the location of our galaxies with those at lower redshift, we derive a linear regression, with the slope fixed to 0.81 i.e. the one for $z=5$ using the formulae of \citet{speagle14}. The regression is derived to be $\log {\rm SFR} = 0.81 \log (M_*/10^{8}\,M_\odot) + 0.31$, with the intrinsic scatter of $0.37$\,dex. 

While the full details of the selected sources will be presented in a forthcoming paper, we highlight two F150W-dropout galaxies ($z\simgt13$). One, JADESGDS-30934, is spectroscopically confirmed to be at $z=13.2$ \citep{curtis-lake23}. The other object, PRIMERCOS-38203, is a newly identified photometric candidate source at $z=13.8_{-1.0}^{+1.1}$ in the PRIMER-COSMOS field (Fig.~\ref{fig:galfit}). Despite the relatively shallow depth in the field, the source exhibits clean non-detection in F090W, F115W, and F150W and high-S/N detection in F200W (S/N=9.2) and in the IR-detection band (S/N=16.1). The observed UV magnitude ($M_{\rm UV}=-20$\,mag) and the derived stellar mass ($\logm=8.0$) are both moderate and comparable to other galaxy candidates at these redshift \citep[e.g.,][]{morishita23,finkelstein23,bouwens23b}. While the F200W-dropout selection covers up to $z\sim18.6$, none is identified at $z>14$ in our selection. The number density estimates of the identified sources will be presented in a forthcoming paper.
{Another potentially interesting source is PRIMERUDS-121885 at $z=10.9_{-0.3}^{+0.3}$, whose stellar mass is $\logm = 10.4_{-0.1}^{+0.1}$. This object has relatively red F356W$-$F444W color ($0.76$\,mag), which implies the presence of old populations. However, we caution that the source is identified in PRIMERUDS, which is relatively shallow among the fields. In addition, at the redshift of the source the F444W flux could also be attributed to strong \hb+\oiii\ emissions, which would lead to a smaller stellar mass.}

{Lastly, we observe moderate concentration of sources at $z\approx7$--$7.6$. This is partially attributed to that strong \hb+\oiii\ emitters are more sensitive to our selection, due to the two medium band filters (F410M and F430M) by their making photometric redshift relatively more constrained. Besides, there is an overdensity of emitters identified in the same redshift range in one of the fields (Daikuhara, in prep.).}

\begin{figure}
\centering
	\includegraphics[width=0.5\textwidth]{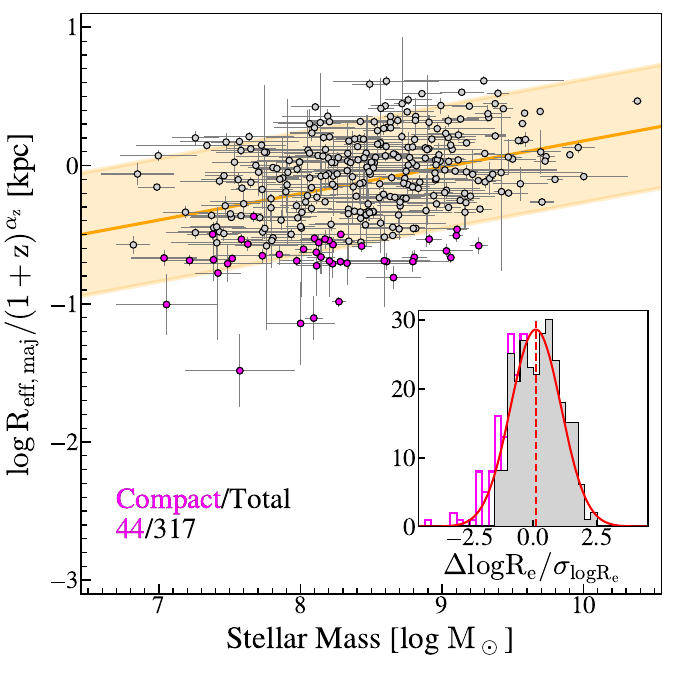}
	\caption{
    Same as in Fig.~\ref{fig:sizemass} but for redshift-corrected size ($=R_{\rm eff, maj}/(1+z)^{\alpha_z}$) of all sources (gray dots) along the determined regression slope (orange solid line, with shades for the $\pm1.5\,\sigma_{\log R_e}$ range). In the inset, the distribution of the normalized size, $\Delta R_{\rm e} / \sigma_{\rm \log R_e}$ is shown, where $\Delta R_{\rm e}$ is the distance from the regression for a given stellar mass. The distribution of the compact sources (open magenta) is shown separately from the remaining extended sources. The Gaussian fit to the extended sources (solid red line) highlights the excess.
	}
\label{fig:normsize}
\end{figure}

\subsection{Size-Stellar Mass Distribution of Galaxies at $5<z<14$}\label{sec:size-mass}

In Fig.~\ref{fig:sizemass}, we show the distribution of galaxies in the size-mass plane for the four redshift ranges. In each redshift panel, we show the size measured in the filter that corresponds to rest-frame $\sim1600\,{\rm \AA}$ i.e. F115W for the F070W-dropout, F150W for the F090W-dropout, F200W for the F115W-dropout, and F277W for the F200W-dropout selection. We adopt the effective radius measured along the major axis, to mitigate the effect by inclination. 

The measured size spans a broad range, $\log R_e\,/\,{\rm kpc} \sim-2$\,--\,$0.3$. Remarkably, at $\logm < 9$, many galaxies are characterized by $R_e<0.3$\,kpc ($<0.\!''07$), which is below the resolution limit afforded by HST/WFC3-IR. Thus, the spatial resolution of NIRCam, $\sim0.11{-}0.14$\,kpc, is essential to study typical star-forming galaxies at these redshift (see also Sec.~\ref{sec:compact}).

We investigate their distribution on the size-mass plane by linear regression analyses. By following \citet[][also \citealt{ferguson04,vanderwel14}]{shen03}, we parameterize size by a log-normal distribution as a function of stellar mass and redshift: 

\begin{equation}\label{eq:sizemass}
\log R_e (M_*, z) / {\rm kpc} = \alpha \log (M_*/M_{\rm 0}) + B(z),
\end{equation}
where we describe the intercept as
\begin{equation}\label{eq:B}
B(z) = \beta_z + \alpha_z \log (1 + z).
\end{equation}
We adopt a pivot mass $M_{\rm 0}=10^{8}\,M_\odot$. The model distribution, $N(\log R_e (M_*), \sigma_{\rm \log R_e})$, prescribes the probability distribution for observing $\log R_e$ for a galaxy with the stellar mass $M_*$ with an intrinsic scatter of $\sigma_{\rm \log R_e}$. By making the intercept $B(z)$ a function of redshift, we are able to evaluate the size distribution with a consistent slope determined by the entire sample. Although the redshift evolution of the slope is of interest, previous studies have reported little changes over a much broader cosmic time range than what is explored in this study \citep[e.g.,][]{vanderwel14,shibuya15}. As we demonstrate below, our findings also reveal no significant evolution in slope from these studies, supporting our employing a single slope across the entire redshift range.

The regression is determined by using {\tt emcee} \citep{foreman14}, with the number of walker is set to $N=50$ and the iteration to $10^4$. We only include those not flagged in structural fitting analysis in Sec~\ref{sec:size}. Measurements uncertainties in size, stellar mass, and redshift are used in the calculation of likelihood. The derived regression is shown in Fig.~\ref{fig:sizemass} along with the measured sizes in four redshift panels. In Fig.~\ref{fig:normsize}, we also show the redshift-corrected size, $R_e / (1+z)^{\alpha_z}$, as this represents the actual variants evaluated in the fitting. We report the determined parameters in Table~\ref{tab:param}. 

The derived slope, $\sim0.20\pm0.03$, is similar to what was found in \citet[][$\sim0.3$]{mosleh11} for Lyman Break galaxies at $z\sim3.5$ and \citet[][0.18--0.25]{vanderwel14} for late-type galaxies at $z\sim0.3$\,--\,$3$, despite the latter being observed in different rest-frame wavelengths (but see the following and Sec~\ref{sec:uvtoopt}). 
The slope of the size-mass relation would reflect the interplay between the intrinsic compactness and concentrated dust attenuation in the core of massive star-forming galaxies at high redshift~\citep{Roper2022}. In fact, negative slopes of the size-mass and size-luminosity relations have been found in cosmological simulations, e.g. BlueTides~\citep{marshall22}, IllustrisTNG~\citep{Popping2022,Costantin2023}, FLARES~\citep{roper22}, and THESAN (Shen et al. 2023, in prep.) simulations. Our finding of a positive slope is consistent with the idea that some massive galaxies in our sample may already possess a moderate amount of dust in its core.

We note that the derived intrinsic scatter, $\sigma_{\log R_e}$, is relatively large ($\sim0.3$) compared to those at lower redshift in the literature ($\sim0.2$). The scatter in size distribution is ought to reflect the initial condition of dark matter halos, such as the distribution of spin parameter \citep[e.g.,][]{bullock01b}. Thus, the observed larger scatter would imply the presence of galaxies that experienced non-linear behavior, such as merger and other dissipative processes, and deviate from the distribution predicted by a simple galactic disc formation model \citep[e.g.,][]{mo98}. In fact, we observed a number of galaxies that are barely resolved in the NIRCam images, some of which are located far below ($>1.5\,\sigma$) the derived regression (Sec.~\ref{sec:compact}). We repeat the regression analysis by excluding these unresolved sources but with the slope fixed to the one derived above. The derived scatter from this regression is $\sim0.25$, moderately reduced from the analysis of the full sample (Table~\ref{tab:param}). 

{Our sample includes both spectroscopic and photometric sources. In particular, the majority ($\sim90\,\%$) of the F070W-d sample are spectroscopic, due to the lack of F070W-filter coverage. To investigate the impact by the spectroscopic sources, we repeat the regression analysis by only using photometric sources. We find a consistent result ($\alpha=0.23\pm0.04, \beta_z=-0.10\pm0.32$, and $\alpha_z=-0.56\pm0.33$), within the uncertainty range. The slight decrease in $\alpha_z$ ($+0.14$, stronger $z$-evolution) can be attributed to the fact that the spectroscopic sources are smaller in size that photometric sources and that they dominate the lowest redshift bin. These differences in the regression result do not change our conclusion.}

\subsection{Redshift Evolution of Galaxy Sizes}\label{sec:zsize}
\begin{figure}
\centering
	\includegraphics[width=0.48\textwidth]{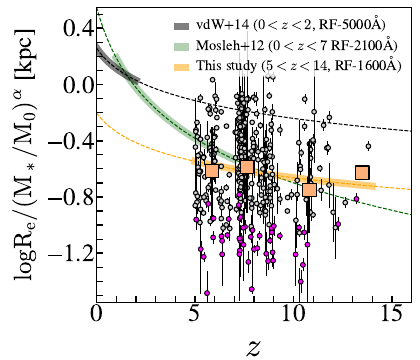}
	\caption{
    Redshift evolution of rest-frame UV size of our galaxies (gray circles). The size of individual galaxies shown here is corrected to the pivot mass (Sec.~\ref{sec:size-mass}). The redshift evolution of the intercept, $\propto (1+z)^{\alpha_z}$, is shown (orange dashed line), along with those derived in two previous studies (green line for \citealt[][]{mosleh12} and black line for \citealt[][]{vanderwel14}). We note that the size trend of both previous studies is also scaled to the same pivot mass. Median sizes (large symbols) are derived in each redshift window. Those identified as compact (Sec.~\ref{sec:compact}) are shown by magenta symbols. 
     }
\label{fig:zsize}
\end{figure}

In Fig.~\ref{fig:zsize}, we show the redshift trend of the UV sizes derived through the regression analysis above. We note that the redshift evolution represent, by design, for the mass-corrected size, $R_e/(M_*/M_0)^{\alpha}$. We find that the derived redshift evolution, $\propto(1+z)^{\alpha_z}$ with $\alpha_z\sim-0.4\pm0.2$, is much less significant than the one derived for rest-frame UV ($\sim2100\,{\rm \AA}$) size of Lyman-break galaxies at $0<z<7$, with $\alpha_z = -1.2 \pm 0.1$ \citep[][also \citealt{oesch10,shibuya15} who found a similar value]{mosleh12}.

The primary cause of the discrepancy can be attributed to the redshift range of the sample. The aforementioned studies derived the redshift evolution by including lower-redshift galaxies ($z\simlt1$ in \citealt[][]{mosleh12,shibuya15} and $2<z<8$ in \citealt{oesch10}). Indeed, \citet{curtis-lake16} found a much slower evolution, $\alpha_z$, for the LBG sample at $4<z<8$, and argued that the discrepancy is partially attributed to a stronger evolution in UV sizes at $z<5$ \citep[see also][who found little size evolution from $z\sim7$ to $6$]{oesch10}. 

To investigate this, we derive the redshift evolution by combining the median sizes presented in Fig.~\ref{fig:zsize} and in \citet[][for $1<z<7$]{mosleh12}, and we do find a stronger evolution ($\alpha_z\sim-1.2$). This supports the idea that the redshift evolution of the average UV size is much slower at $z>5$ than lower redshifts. However, we note that the outcomes is likely subject to a range of systematic factors, such as the weighting of each size measurement, the inclusion or exclusion of specific data points, and potential sample mismatches. 

In Fig.~\ref{fig:zsize}, we also show the evolution of rest-frame optical sizes derived for low-mass ($\logm\sim9$) late-type galaxies at $0<z<2$ \citep{vanderwel14}. Interestingly, 
the extrapolated sizes from the fit to our redshift range exhibit a significant offset, even after accounting for the mass difference between the two samples. Although the offset might be attributed to differences in the bands used for size measurements (i.e. rest-frame $5000\,{\rm \AA}$ in \citealt{vanderwel14}), we find that our galaxies, on average, do not show a significant offset between the two bands. This reinforces our earlier interpretation that size evolution could be more pronounced at $z\simlt5$, primarily driven by the build-up of massive bulges and/or outskirts, which would enhance color gradients, as seen at lower redshifts. We discuss this in more detail in Sec.~\ref{sec:uvtoopt}.

\subsection{Identifying Blue Compact Sources}\label{sec:compact}
In our size analysis above, we have identified a number of compact sources that are near the resolution limit of NIRCam. Previous studies using NIRCam data also reported a few of such compact sources \citep{yang22,castellano22b,naidu22,ono22}. 
We define those that satisfy either of the following as {\it compact}: 
\begin{itemize}
    \item[] $i$)~$R_e < {\rm FWHM}/2$ in the measured band; or
    \item[] $ii$)~$\Delta {R_e} < -1.5\,\sigma_{\log R_e}$
\end{itemize}
where $\Delta R$ is the difference of the measured size from the inferred size by the linear regression for the stellar mass and redshift of the source. Note that $R_e$ used here is apparent size. 

With the criteria, we find \ncom\ compact sources from our sample ($\sim13\,\%$), including \ncomspec\ that are spectroscopically confirmed. We then exclude five of the classified compact sources, whose SED is better fitted with a brown dwarf template over galaxy templates (Sec.~\ref{sec:sel}). The classified compact sources are marked in Fig.~\ref{fig:sizemass}. Individual cutout images are presented in Appendix~\ref{appen:compact}.

In Fig.~\ref{fig:normsize}, we show the distribution of the normalized size, $\Delta R_e / \sigma_{\log R_e}$. The distribution clearly shows an excess at smaller size when compared to the distribution of the remaining non-compact sources. We note that the fraction of the identified compact sources is well above the number expected at $1.5\,\sigma$ for a normal distribution (i.e. $6.7\,\%$). The compact sources follow a similar distribution as other extended sources in physical properties, such as stellar mass, star formation rate (see also Fig.~\ref{fig:sfms}), $\beta_{\rm UV}$, except for star formation surface density (Sec.~\ref{sec:sfrd}). The ISM properties of the compact sources are further investigated in Sec.~\ref{sec:msa}.

\section{Discussion}\label{sec:disc}

\begin{figure*}
\centering
	\includegraphics[width=0.49\textwidth]{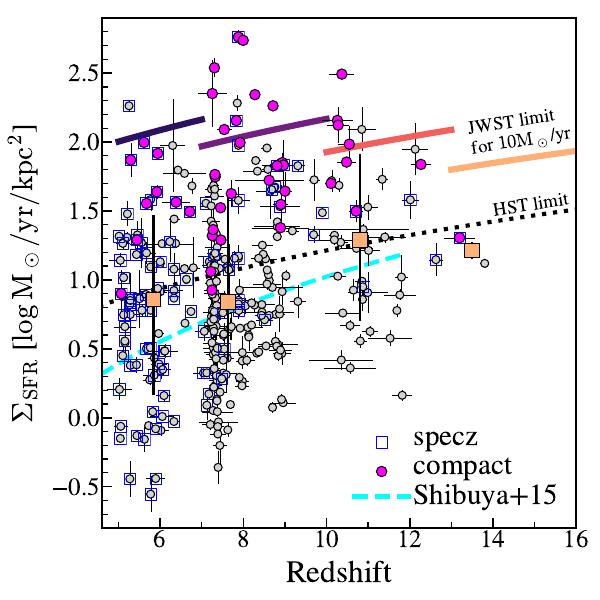}
	\includegraphics[width=0.49\textwidth]{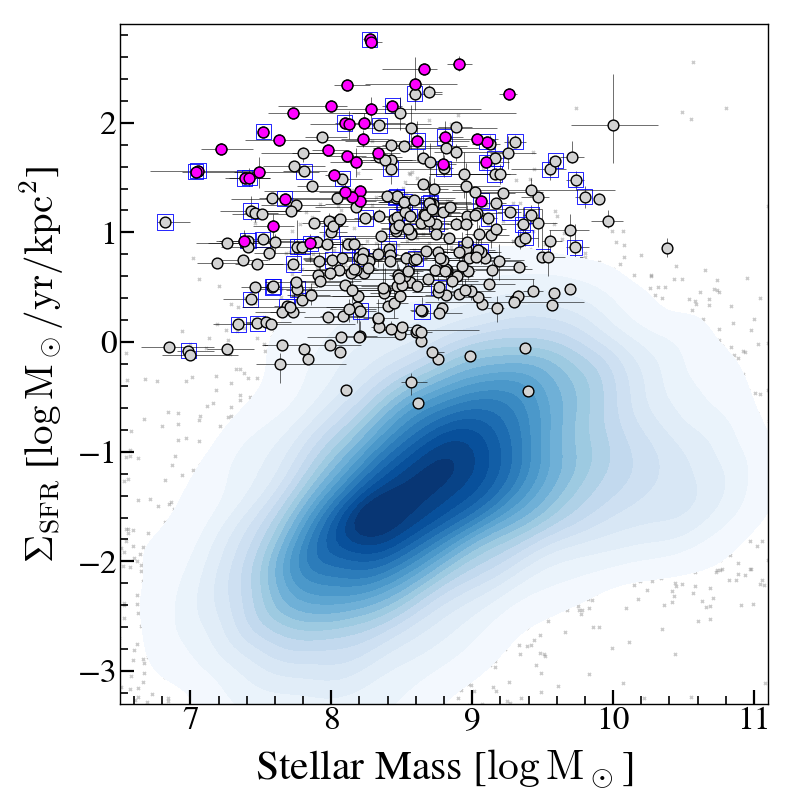}
	\caption{
 ($Left$): Star formation surface density ($\Sigma_{\rm SFR}$) of our final sample (gray circles) as a function of redshift. The median values at each dropout redshift window are shown by larger symbols (squares). The upper limits for $\Sigma_{\rm SFR}$ by spatial resolution limit, when assuming SFR of $10\,M_\odot\,/\,{\rm yr}$, are shown for the corresponding JWST NIRCam filters (solid curved lines) and the HST WFC3-IR F160W (black dotted line). The fit derived for LBGs in HST data \citep[][cyan dashed line]{shibuya15} is shown. 
 ($Right$): $\Sigma_{\rm SFR}$ as a function of stellar mass. In the background, we show the density contour for the distribution of galaxies at $0.3<z<3.5$, taken from the 3DHST catalog \citep{skelton14,vanderwel14}. Those outside the lowest contour level are shown individually (crosses). Only $\sim6\,\%$ of the 3DHST sample (and $<0.01\,\%$ of the $z\sim0$ Sloan sample) is found at $\Sigma_{\rm SFR}>1\,{\rm M_\odot / yr / kpc^2}$; the majority of our sample is located above the value.
	}
\label{fig:sfrd}
\end{figure*}

\begin{figure}
\centering
	\includegraphics[width=0.45\textwidth]{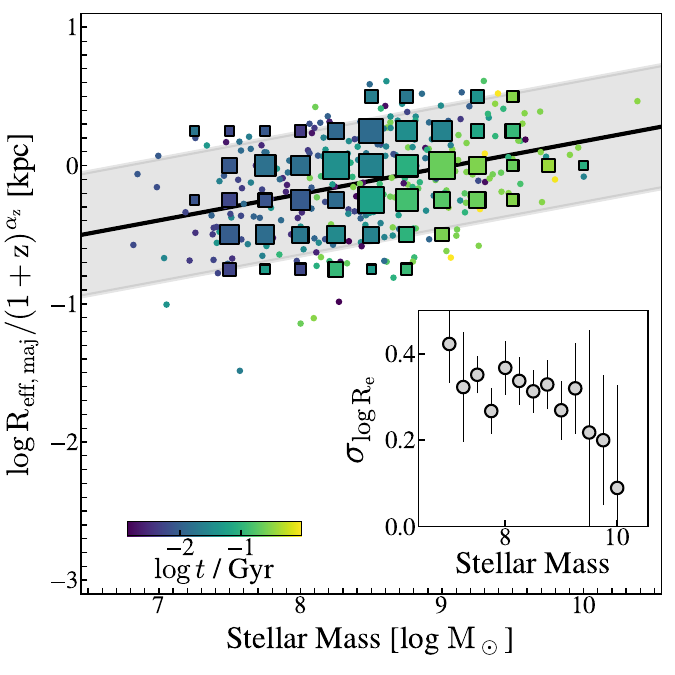}
	\caption{
    Same as in Fig.~\ref{fig:normsize} but symbols are color-coded by age ($\log t$) derived in our SED analysis. The median age in each grid is shown by larger symbols (squares), with the symbol size scaled by the number of galaxies in the grid. In the inset, the scatters in size measured at each mass bin are shown. The error of each scatter measurement is estimated by bootstrapping. Clearly, the radius and age of the stellar population increase, while the scatter in sizes decreases, with increasing stellar mass.
	}
\label{fig:normsize_logt}
\end{figure}

\subsection{High Star Formation Efficiency Revealed by NIRCam Imaging}\label{sec:sfrd}
Star formation (rate) surface density,  
\begin{equation}    
\Sigma_{\rm SFR}\,/\,{\rm M_\odot\,yr^{-1}\,kpc^{-2}} = {\rm 0.5\,SFR}\,/\,\pi R_{e}^2,
\end{equation}
is known as an excellent proxy for inferring the current mode of star formation. In Fig.~\ref{fig:sfrd}, we show the distribution of $\Sigma_{\rm SFR}$ of our sample. Overall, the median values of our galaxies are consistent with previous studies with HST \citep{oesch10,shibuya15}.

However, we observe a large scatter along the vertical axis. In fact, we find a large fraction of galaxies of $\log \Sigma_{\rm SFR}\simgt1.5$, comparable to local ULIRG/starburst \citep{scoville00,dopita02,kennicutt12} and high-$z$ sub-millimetre galaxies \citep[e.g.,][]{daddi10}. Such high-$\Sigma_{\rm SFR}$ galaxies were not reported in previous HST studies at similar redshift \citep{oesch10,ono13,holwerda15}, except for a few cases in cluster lensing fields \citep[e.g.,][]{kawamata15,bouwens22}. Obviously, $\Sigma_{\rm SFR}$ estimates are limited by imaging resolution, and thus the previous estimate for smaller galaxies identified in HST data often remained as lower limits \citep[e.g.,][]{morishita21,fujimoto22,ishikawa22}. 

The observations in Fig.~\ref{fig:sfrd} also demonstrate the potential for JWST NIRCam observations to probe star formation activity at low stellar masses ($M\lesssim 10^8 M_{\odot}$). This had been so far limited to indirect probes, such as studies based on follow-up of Gamma Ray Burst afterglows to quantify the fraction of detected host galaxies (e.g, \citealt{Tanvir2012,Trenti2012GRB,McGuire2016}). Those studies hinted the existence of the population of sources now directly probed by NIRCam.  

\subsubsection{Implications for Star-formation Efficiency}
The compact sources identified in Sec.~\ref{sec:compact} dominate the upper range, $\log \Sigma_{\rm SFR}\simgt1.5$. We note that this does not stem from their star formation rates, which are, on average, comparable to those of more extended sources. Instead, it is attributed to their compact nature. These compact sources' physical characteristics are particularly intriguing, as negative feedback is likely more effective within confined systems. The observed high values thus imply efficient gas fueling within these compact sources, potentially facilitated by processes such as the loss of angular momentum through mergers (see also Sec.~\ref{sec:msa}). 

In addition, the mass-dependence of $\Sigma_{\rm SFR}$ could hint at the efficiency of star formation, as the system's mass is tightly linked to the regulation of star formation. In the right panel of Fig.~\ref{fig:sfrd}, we show the distribution of $\Sigma_{\rm SFR}$ as a function of stellar mass. Of particular interest is the high mass range, $\sim10^9\,M_\odot$. 
In this mass range, the shocked gas remains hot \citep[e.g.,][]{birnboim03,dekel06,stern21}, resulting in a reduced star formation efficiency, as seen at lower redshifts. The observed high values for our sample thus imply that our sources still hold a high efficiency within this mass regime. We calculate the median mass-doubling time by $t=M_*/{\rm SFR}$ and find $\sim15$--$90$\,Myr for our sample. These considerably small mass-doubling times suggest that some galaxies in our sample could evolve to $\logm \sim 11$ by $z\sim5$, provided that the efficiency remains similar in the following $\sim0.5$--1\,Gyr. 

\subsubsection{Implications for the Growth of Galaxies}
At these early cosmic time,
comparing the radius-mass relationship of galaxies as a function of the age of the stellar population can
reveal how galaxies grow (Fig.~\ref{fig:normsize_logt}). In hierarchical structure formation, galaxies grow through the
merger of smaller mass galaxies. The merger process results in the infall of stars but also allows new
star-formation to occur through shocking of the infalling gas. If galaxies grow predominantly by cold
gas accretion \citep[e.g.][]{dekel06}, one would expect their size to evolve predominantly through secular evolution which would
be driven by dynamical friction between the stars and the accreted gas i.e. galaxies would become
smaller as they built up their stellar mass. We find two clear trends in our sample of high-$z$ galaxies:

\begin{itemize}
    \item[1.] Both radius and age increase with stellar mass. High mass galaxies harbor older ($\sim$\,several 100 Myr) stellar populations compared to lower mass galaxies, which harbor populations of age $\sim$10-100 Myr.

    \item[2.] The scatter in radii at fixed mass is larger by 0.2\,dex for low mass galaxies than for high mass galaxies.
\end{itemize}

Since there is no selection effect that prevents the selection of small, high mass galaxies, the implication
of these two observational trends is that galaxies form inside out. Small, low-mass galaxies have a large
scatter in their sizes likely due to a combination of gas accretion and merger events. As they form stars, the 
stars get scattered due to three-body interactions, resulting in a growth in size and stellar mass.
That is consistent both with the size evolution with cosmic time and the radius and stellar age evolution with mass.
Although the sensitivity of the data at the present time are not adequate to constrain minor merger rates
at these redshifts, future deeper surveys will help develop this hypothesis further.

\subsubsection{Comparison with Local Lyman-Break Analogs}
It is illustrative to compare the luminosity surface density of these galaxies with similar objects in the local Universe \citep{hoopes2007, overzier09, Shim2013}. Although objects with such high surface densities of star-formation exist at $z\sim0$, less than 0.01\% of galaxies in the Sloan sample have $\Sigma_{\rm SFR}>1$\,${\rm M_\odot\,yr^{-1}\,kpc^{-2}}$. Even at higher redshifts, the fraction remains small \citep[$\sim6\,\%$ at $0.3<z<3.5$;][]{skelton14}.
In contrast, the fraction is $\sim$100\% in our sample. The $z\sim0$ objects have a median mass
of 10$^{8.9}$\,M$_{\odot}$ and span the full range of metallicity from sub-solar to super-solar.
They are not particularly biased towards young ages as determined from the strength of the Balmer break. Morphologically,
we see some evidence of mergers in the high-$z$ sample with $\sim 20\, \%$ of galaxies having nearby companions (besides the \nfailed\ galaxies flagged in the size analysis). Taken together, the implication therefore is that the high surface densities are gas-rich galaxies with the gas being relatively concentrated towards the nucleus of the galaxy indicating that the late stages of the merging process is driving the inflow of gas towards the nucleus of the galaxies.

\begin{figure*}
\centering
	\includegraphics[width=0.49\textwidth]{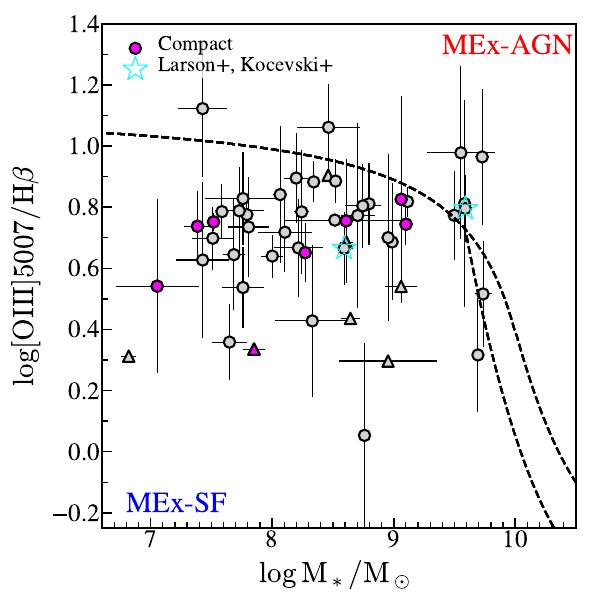}
	\includegraphics[width=0.47\textwidth]{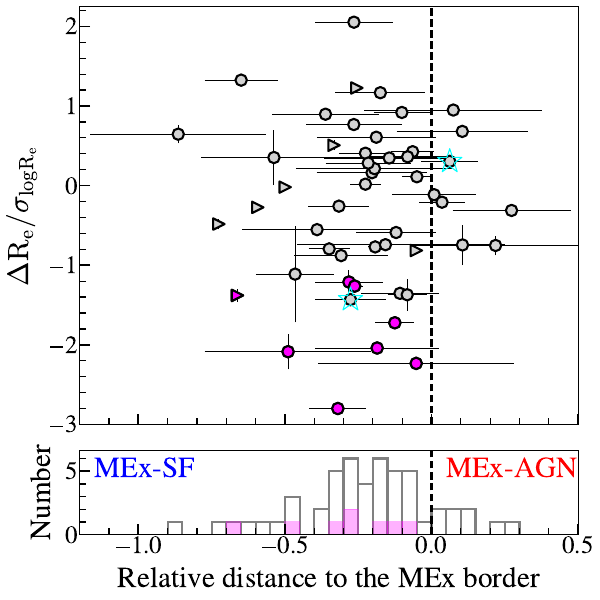}
	\caption{
    ({\it Left}):
    Distribution of \nmex\ galaxies at $5<z<9.5$ that have robust \oiii$_{\lambda5007}$/\hb\ measurements by NIRSpec MSA, in the mass-excitation (MEx) diagram (circles). For those with \hb\ detected at S/N\,$<3$, lower limits for the ratio are shown (triangles). Compact galaxies (as defined in Sec.~\ref{sec:compact}; magenta) and two sources from \citet[][cyan stars]{larson23,kocevski23} are marked separately. The solid lines are shown to indicate the regions dominated by star forming galaxies (left bottom) and AGN (right top) of $z\sim2$ \citep{juneau14}.
    ({\it Right}):
    Distribution of the normalized size ($\Delta R_e/\sigma_{\log R_e}$, an indicator that is used to define compactness in Sec.~\ref{sec:compact}) as a function of the distance to the MEx border separation line. 
    }
\label{fig:mex}
\end{figure*}

\subsection{Nature of the Compact Sources --- NIRSpec Spectroscopic Analysis}\label{sec:msa}
We here investigate the nature of the compact sources identified in Sec.~\ref{sec:size-mass}, specifically aiming to observe if they exhibit any evidence of AGN. Traditionally, point-source-like morphologies of UV bright sources have been considered as evidence of AGN. However, it has been shown that the compactness does not always indicate the presence of AGN \citep[e.g.,][]{morishita20}, and vice versa \citep[see, e.g.,][for faint AGN with extended features]{matthee23}. As such, a comprehensive approach is required to answer the question.

A subset of our sample galaxies have spectroscopic coverage by NIRSpec/MSA, taken as part of CEERS, GLASS-ERS, and JADES. 
Following \citet{morishita23b}, we reduce the MSA spectra, using {\tt msaexp}\footnote{https://github.com/gbrammer/msaexp}. For the extracted 1d spectrum of each source, we fit the line profiles of \hb\ and \oiii-doublets with a Gaussian, after subtracting the underlying continuum spectrum inferred by {\tt gsf} in Sec.~\ref{sec:sed}. For each line, the total flux of each line is estimated by integrating the flux over the wavelength range of 2$\times$FWHM derived from the gaussian fit. In the following analysis, we include sources with measured line S/Ns above 3 for the \oiii$_{\lambda5007}$ line (N=\nmex); when \hb\ is not detected above the same significance, we quote the $3\,\sigma$ flux limit measured at the wavelength of the line over the same line width derived for \oiii$_{\lambda5007}$. 

In Fig.~\ref{fig:mex}, we show the sample in the mass-excitation (MEx) diagram, a conventional diagnostic for AGN and star-forming galaxies \citep{juneau11}. Among the \nmex\ objects, we find eight sources located within the MEx AGN region defined by \citet{juneau14}. One of the sources, CEERS7-18822, was previously reported to have a tentative ($\sim2.5\,\sigma$) broad component in \hb\ \citep[][]{larson23}, agreeing with the classification here. CEERS7-18822 exhibits extended structures in F115W, which confirms our classification of this source not to be compact. 

None of our compact sources is classified as MEx-AGN. While JADESGDS-18784 is located near the MEx border, with $\log\,$\oiii/\hb\,$=0.83\pm0.34$ (classified as MEx-AGN within the uncertainty range), we do not confirm any features that immediately supports the presence of AGN (i.e. broad line features or high-ionization lines). None of the other compact sources show AGN signatures either, except for CEERS6-7832 which was reported to have a broad ($\sim2000$\,km/s) component in its \ha\ line \citep[][as CEERS\_1670\footnote{The other object reported by \citet{kocevski23}, CEERS\_3210, does not pass our $S/N>8$ selection criterion and is not included in our final sample.}]{kocevski23}. 

However, this does not completely rule out the absence of AGN. Firstly, the discriminating power of the MEx diagram could be lower near the transition region. As shown by \citet{juneau14}, for the range of $\log$\,\oiii/\hb\ line ratios probed here ($\simgt0.4$) there could be 10--30\,$\%$ of AGN present even inside the MEx-SF region at $\logm\sim10$. For example, the aforementioned CEERS\_7832 is found in the MEx-SF region. The accuracy of the AGN classification at a lower mass range is not known due to the lack of data in the previous study. We also note that the observed ratio for our sample ($\log$\,\oiii/\hb\,$\simgt0.6$) is relatively high, compared to the star-forming galaxies of a comparable mass in \citet[][also \citealt{shapley15}]{juneau14}. Such a high ratio can still be achieved by a stellar-only configuration, but requires, e.g., high electron density \citep[][]{reddy23}. 

We note that, given the evolving ISM properties at these redshifts, there is likely a shift of the MEx boundary toward a higher line ratio, as is the case for $z\sim0$ to $z\sim2$. As such, it is still possible that any of our MEx-SF sources near the border may host an AGN, if not a broad line AGN. 
Furthermore, as has been demonstrated in the local Universe \citep{Ho97}, it is possible to bury a low-luminosity Seyfert or LINER-type nucleus in a galaxy without detecting a component of broad line emission.
Vice versa, some lower-mass MEx-AGN sources here could turn to be MEx-SF due to the potential redshift evolution of the boundary.

Nonetheless, the absence of clear AGN evidence implies that the observed high ratios are driven by high surface density of star formation. When we fit high values of \oiii/\hb\,$>3$ with radiative shock models \citep[e.g., MAPPING III;][]{allen08}, we find that shock velocities of a median of $500$\,km/s are required. Ratios of $\sim$10 can only be achieved with high shock velocities which could be correlated with high surface density of star-formation (or strong AGN activity). Achieving the observed high star formation surface density ($>100\,M_\odot / {\rm yr / kpc^2}$) is considered challenging, due to the presence of negative feedback. Such high density is expected only in an extreme environment, where an abundant of gas is available, and/or in (post-)merging systems where gas could rapidly fall in. By comparing with a numerical simulation, \citet{ono22} found that such a compact galaxy is in a temporary compact star-forming phase triggered by recent major mergers. \citet[][]{roper22} found a large fraction of blue compact ($\sim100$--300\,pc) galaxies {\it and} these galaxies to have little contribution by AGN in the FLARES simulations \citep[also,][Shen, X., in prep.]{marshall22}. 
Future work will compare derived physical parameters from the emission lines, with the measured star-formation rate density.

\begin{figure*}
\centering
	\includegraphics[width=0.94\textwidth]{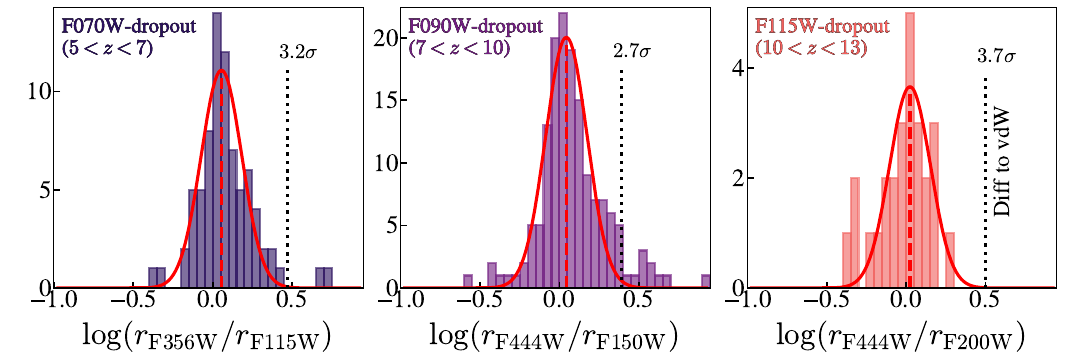}
	\caption{
    Distributions of rest-frame UV-to-optical size ratio for three redshift bins. The first two panels compares rest-frame UV to optical ($\sim5000\,{\rm \AA}$) sizes, whereas the last panel at a shorter wavelength ($\sim3700\,{\rm \AA}$) due to the filter availability. Only resolved galaxies in the two filters are included. The F150W-dropouts are not shown, as neither of their size is resolved in F444W. Each distribution is fit with Gaussian (red lines), with the mean position indicated by vertical dashed lines. The difference to the extrapolated size in rest-frame optical of \citet{vanderwel14} (observed in Fig.~\ref{fig:zsize}) is shown in each panel (dotted vertical lines); the difference is measured to be 3.2, 2.7, and 3.7\,$\sigma$, respectively. 
 	}
\label{fig:size_filts}
\end{figure*}

\subsection{Comparison of sizes in rest-frame UV and optical}\label{sec:uvtoopt}
In this study, we have analyzed the sizes of galaxies at a rest-frame wavelength of $\sim1600\,{\rm \AA}$. In Sec.~\ref{sec:size-mass}, we found that the average sizes of our galaxies are much smaller than those predicted from the extrapolation of \citet{vanderwel14} at the corresponding redshift. A possible explanation for the discrepancy could be the rest-frame wavelengths where the size is measured in the previous study (i.e. $\sim5000\,{\rm \AA}$). We investigate this by repeating our size analysis but in a filter that corresponds to $\sim5000\,{\rm \AA}$, i.e. F356W for the F070W-dropouts, and F444W for the F090W-dropouts. For the other two ranges at higher redshift, we use the reddest filter available (F444W), which corresponds to $\sim4000\,{\rm \AA}$ and $\sim3000\,{\rm \AA}$, respectively. In Fig.~\ref{fig:size_filts}, we show the distribution of the size difference in these two wavelengths. We find that the difference of size in the two filters is negligible on average, and thus conclude that the wavelength difference cannot explain the offset seen in the extrapolated size of \citet{vanderwel14} observed in Fig.~\ref{fig:zsize}. Instead, we speculate that the extrapolation of their relation may not persist in the redshift range far beyond their probed redshift range $z<3$. 


In fact, we have seen in Sec.~\ref{sec:size-mass} that the size evolution is much slower ($\alpha_z=-0.4$) than the one found in previous studies of rest-frame UV size \citep[$\sim-1.2$;][]{mosleh12,shibuya15}. While a comprehensive analysis covering a wide redshift range would be necessary, we attribute the observed conflict to the build-up of complicated structures in galaxies, such as a central massive bulge and young star forming disk. At lower redshifts, we observe a more pronounced difference in sizes between different wavelengths, driven by radial color gradients within the systems \citep[e.g.,][]{vulcani14}. Similarly, \citet{shibuya15} found that at $z=1.2{-}2.1$ the average UV size is smaller than the optical size by $\sim20\%$ in the low-mass regime ($\sim10^9\,M_\odot$), while the trend is reversed at the high-mass end ($\simgt10^{11}\,M_\odot$; also see \citealt{szomoru12,vanderwel14}). 

On the other hand, the resemblance of galaxy UV and optical sizes of high-redshift galaxies is not unexpected given the rapid assembly of the stellar content \citep[also see][]{yang22b,treu23}. For our galaxies, we have estimated the mass-doubling time to be $\simlt100$\,Myr in Sec.~\ref{sec:sfrd}. This time scale is comparable to or even smaller than the star formation time scale that the UV tracer is sensitive to~\citep{Murphy2011,Flores2021}. Consequently, the stellar content detected by the UV will predominate over the total content integrated over the full star formation history, which is probed by observations at optical wavelengths. The majority of our galaxies are actively building their structures inside-out, developing rapidly enough to maintain coherence. 

{Lastly, we investigate the rest-frame optical size of the compact sources identified in this work. Of \ncom, we find 13 ($\sim29\,\%$) show resolved morphology in optical filters. This is opposite from the general trend discovered above, and implies that some fraction of these compact sources are likely experiencing a secondary burst, in a relatively small area, after the initial build-up of stellar structure. This is in particular relevant to the stochastic nature of star formation in early galaxies, which may have non-negligible effect of the observed UV magnitude measurements and consequently UV luminosity functions. 
}

\section{Summary}\label{sec:sum}
In this study, we have identified \nsrc\ galaxies at $5<z<14$ in legacy fields of JWST and analysed their rest-frame UV and optical sizes through JWST NIRCam images. The imaging data used here were collected from several public programs in Cycle~1, resulting in a combined effective area of 358\,arcmin$^2$. With a robust ($8\,\sigma$) selection of \nsrc\ galaxies, made possible by the unprecedented area coverage provided by JWST, we have conducted the first systematic exploration of the size-mass relation of galaxies in the first billion years. The key findings are as follows:
\begin{itemize}

    \item The slope of the size-mass relation was derived via linear regression analyses and found to be $\alpha\sim0.2$, similar to those of star forming galaxies at $z<3$ but scaled down in size by 0.4 dex. The derived intercept was found to evolve with $\propto(1+z)^{-0.4}$, much slower evolution than those found in previous studies of rest-frame UV sizes of galaxies at lower redshifts.

    \item 
    By using the results from our linear regression analysis, we identified \ncom\ compact sources that are marginally resolved in NIRCam imaging. These compact sources account for $\sim13\,\%$ of the full sample presented here.

    \item We found that our sources overall have high star formation surface density ($\simgt1\,M_\odot {\rm yr^{-1}\,kpc^{-2}}$), with the newly identified compact sources being as high as $\sim300\,M_\odot {\rm yr^{-1}\,kpc^{-2}}$. We demonstrated that the absence of a clear declining trend indicates that star formation efficiency may remain high even at the high mass range; if the observed high efficiency remains similar in the following $\sim0.5$--1\,Gyr, some of our sources would evolve to $\sim10^{11}\,M_\odot$ by $z\sim5$.

    \item For \nmex\ sources with available NIRSpec/MSA data, we investigated their ISM via the \oiii-to-\hb\ line ratio. None of the compact sources are confidently classified as AGN in the Mass-Excitation diagram; however, the nature of the compact sources remains to be conclusively elucidated in a future study. A potential explanation for the observed high line ratios is high shock velocities, driven by intense star-formation characterized by high $\Sigma_{\rm SFR}$.
    
    \item We found that the sizes in rest-frame and optical wavelengths are on average consistent. We attributed this to the short mass-doubling time (i.e. $1/{\rm sSFR}<100$\,Myr) of our sources, implying that they are actively building their structure coherently, and are thus dominated by young stars. 
    
\end{itemize}

With the unprecedented resolution and sensitivity provided by JWST, this work demonstrated a comprehensive size analysis of galaxies in the first billion years. Of particular remarks are the newly discovered compact population. Specifically, the physical mechanisms that maintain the observed high star formation rate in such compact systems, which are likely under the strong influence by negative feedback, remains an open question. Recent JWST observations have identified a number of faint AGN \citep{onoue23,matthee23} and a complex of dusty AGN + young stellar populations \citep{furtak23,labbe23b,akins23}, suggesting a greater prevalence of AGN in high-$z$ galaxies than previously thought. These emerging findings raise a caveat that the estimated {\it star formation} rate from our SED analysis may not accurately represent the intrinsic value, even though our spectroscopic analysis on the subsample did not find any immediate signatures of AGN. Future spectroscopic followups of the compact sources will provide further insight into their nature and their potential impact in a broad cosmological context.

\tabletypesize{\footnotesize} \tabcolsep=0.11cm
\begin{deluxetable*}
{@{\extracolsep{4pt}}rrrrrrrrrr@{}} 
    \tablecolumns{10}
    \tablewidth{0pt}
    \tablecaption{Linear Regression Best-Fit Coefficients}
    \tablehead{
    \multicolumn{5}{c}{$M_{\rm UV}$--Size relation} & \multicolumn{5}{c}{$M_{\rm UV}$--Stellar mass relation} \\
    \cline{1-5}
    \cline{6-10}
    \colhead{$N$} & \colhead{$\alpha$} &
    \colhead{$\beta_z$} & \colhead{$\alpha_z$} & \colhead{$\sigma_\mathrm{\log R_e}$} &  \colhead{$N$} & \colhead{$\alpha$} & \colhead{$\beta_c$} & \colhead{$\alpha_c$} & \colhead{$\sigma_\mathrm{\log M_*}$}
    }
\startdata
    317  & $-0.09_{-0.02}^{+0.02}$ & $0.08_{-0.21}^{+0.21}$ & $-0.60_{-0.22}^{+0.23}$ & $0.31_{-0.01}^{+0.01}$ & 317  & $-0.28_{-0.03}^{+0.03}$ & $8.83_{-0.04}^{+0.04}$ & $0.76_{-0.08}^{+0.08}$ & $0.47_{-0.02}^{+0.02}$
\enddata
\tablecomments{
Coefficients determined in the linear regression analysis in Appendix~A.
}\label{tab:param_Muv}
\end{deluxetable*}

\section*{Acknowledgements}
We would like to thank the teams of the JWST observation programs \#\,1063, 1180, 1210, 1345, 1837, 2079, and 2738, which a large part of this study is based on, for their countless efforts in carefully designing and planning their programs and their generous consideration of making their valuable data publicly available immediately after the observations or before the expiration of the original proprietary periods. Some/all of the data presented in this paper were obtained from the Mikulski Archive for Space Telescopes (MAST) at the Space Telescope Science Institute. The specific observations analyzed can be accessed via \dataset[10.17909/q8cd-2q22]{https://doi.org/10.17909/q8cd-2q22}.
We thank Zhaoran Liu for kindly providing a spectroscopic catalog compiled from literature studies in the CEERS field. We acknowledge support for this work under NASA grant 80NSSC22K1294. 
The data presented in this paper were obtained from the Mikulski Archive for Space Telescopes (MAST) at the Space Telescope Science Institute. 

{
{\it Software:} 
Astropy \citep{astropy13,astropy18,astropy22}, bbpn \citep{bbpn}, EAzY \citep{brammer08}, EMCEE \citep{foreman13}, gsf \citep{morishita19}, numpy \citep{numpy}, python-fsps \citep{foreman14}.
}



\begin{figure*}
\centering
	\includegraphics[width=0.88\textwidth]{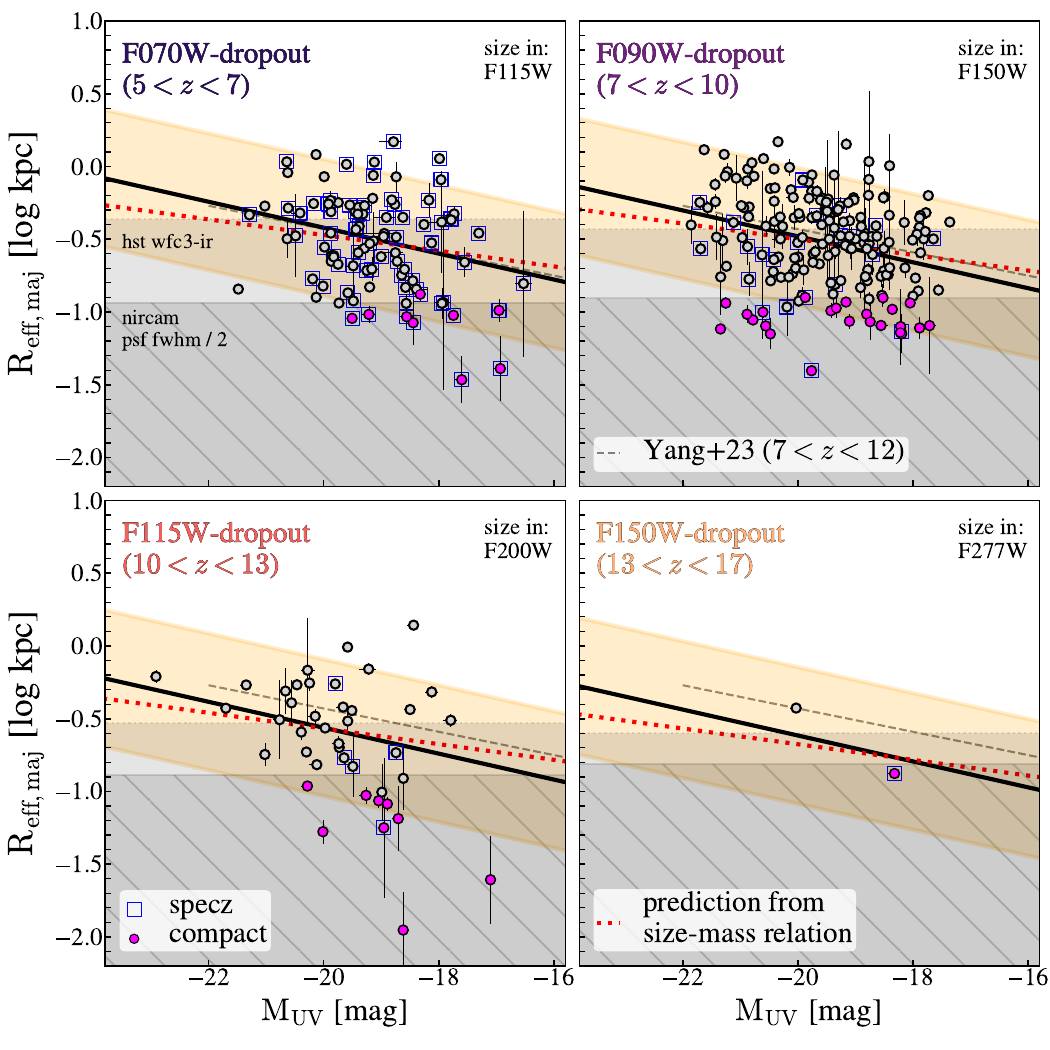}
	\caption{
    Same as Fig.~\ref{fig:sizemass} but the distribution in the absolute UV magnitude ($M_{\rm UV}$)-size plane. For comparison, the slope fit to LBGs at $7<z<12$ \citep[][black dashed lines]{yang22b} is shown.
    The predicted slope from our size-stellar mass relation (Eq.~\ref{eq:sizemass}) by using the $M_{\rm UV}$--stellar mass conversion (Eq.~\ref{eq:muvmass}) is shown (red dashed lines). 
	}
\label{fig:sizeMuv}
\end{figure*}

\begin{figure}
\centering
	\includegraphics[width=0.4\textwidth]{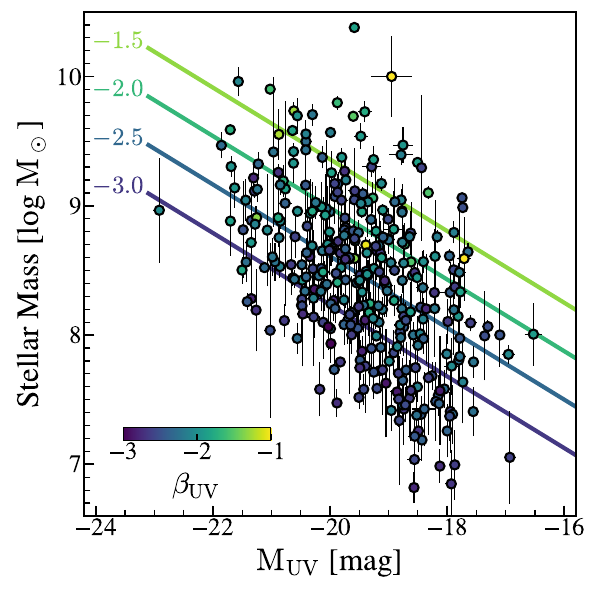}
	\caption{
    Distribution of our sample in the absolute UV magnitude--stellar mass plane (circles). Symbols are color-coded by the value of $\beta_{\rm UV}$. The regression slopes at various $\beta_{\rm UV}$ values, determined using Eqs.~\ref{eq:muvmass} and \ref{eq:muvmassb}, are shown (solid lines).
	}
\label{fig:Muvmass}
\end{figure}

\section*{Appendix~A: Comparison with the $M_{\rm UV}$-Size relation}\label{appen:sizeMuv}
In Fig.~\ref{fig:sizeMuv}, we show the distribution of our sample galaxies in the absolute UV magnitude ($M_{\rm UV}$)-size plane. We follow the regression analysis in Sec.~\ref{sec:size-mass} using the following formulae:
\begin{equation}\label{eq:muvsize}
\log R_e (M_{\rm UV},z) = \alpha (M_{\rm UV} - M_{\rm UV, 0}) + B(z),
\end{equation}
\begin{equation}\label{eq:muvsizeb}
B(z) = \beta_z + \alpha_z \log (1 + z).
\end{equation}
We adopt the pivot magnitude $M_{\rm UV,0}=-20$\,mag. The regression analysis gives a similar slope ($\alpha=-0.09\pm0.02$) to those found in the literature at $z\sim5$ \citep[$-0.1$;][]{huang13} and at similar redshift \citep[$-0.11$--$-0.15$;][]{shibuya15,kawamata18}. The intercept is very similar to \citet{yang22b}, who derived the intercept for $7<z<12$ galaxies using a fixed slope of the \citet{huang13}'s value. The consistent intercept supports that our size measurement is comparable to the literature studies \citep[see also][who compare their sizes with \citealt{yang22b}]{ono22}. The derived parameters are reported in Table~\ref{tab:param_Muv}.

In Fig.~\ref{fig:Muvmass}, we show the distribution of our final sample in the $M_{\rm UV}$-$\log M_*$ plane. The distribution shows a weak trend with $\beta_{\rm UV}$ but not with redshift. We thus characterize the distribution by:
\begin{equation}\label{eq:muvmass}
\log M_*(M_{\rm UV}, \beta_{\rm UV}) = \alpha (M_{\rm UV} - M_{\rm UV, 0}) + B(\beta_{\rm UV}),
\end{equation}
\begin{equation}\label{eq:muvmassb}
B(\beta_{\rm UV}) = \beta_c + \alpha_c (\beta_{\rm UV} - \beta_{\rm UV, 0})
\end{equation}
with the same pivot magnitude $M_{\rm UV,0}=-20$\,mag as above, and $\beta_{\rm UV, 0}=-2$. The regression slope is determined but with a relatively large scatter ($\sigma\sim0.47$)
By using this conversion and the size-stellar mass relation in the main text, we can predict the size-$M_{\rm UV}$ relation (and vice versa). In Fig.~\ref{fig:sizeMuv}, along with the regression derived above, we show the predicted size-$M_{\rm UV}$ relation for the sample, by using the median $\beta_{\rm UV}$ value, in each redshift window. The predicted slope is shallower but overall in agreement with the one derived above within the uncertainty. We provide the determined parameters in Table~\ref{tab:param_Muv}.


\section*{Appendix~B: Physical Properties of the Final Sample}\label{appen:phys}
We report the physical properties of the final \nsrc\ sources at $z>5$ in Table~\ref{tab:prop}. The table contains coordinates, apparent magnitudes (Sec.~\ref{sec:photom}), spectroscopic or photometric redshifts (Sec.~\ref{sec:photz}), size measurements and flags (Sec.~\ref{sec:size} and \ref{sec:compact}), and SED parameters (Sec.~\ref{sec:sed}).

\section*{Appendix~C: On the sample completeness and the robustness of our size measurements}\label{appen:robust}
\begin{figure}
\centering
	\includegraphics[width=0.45\textwidth]{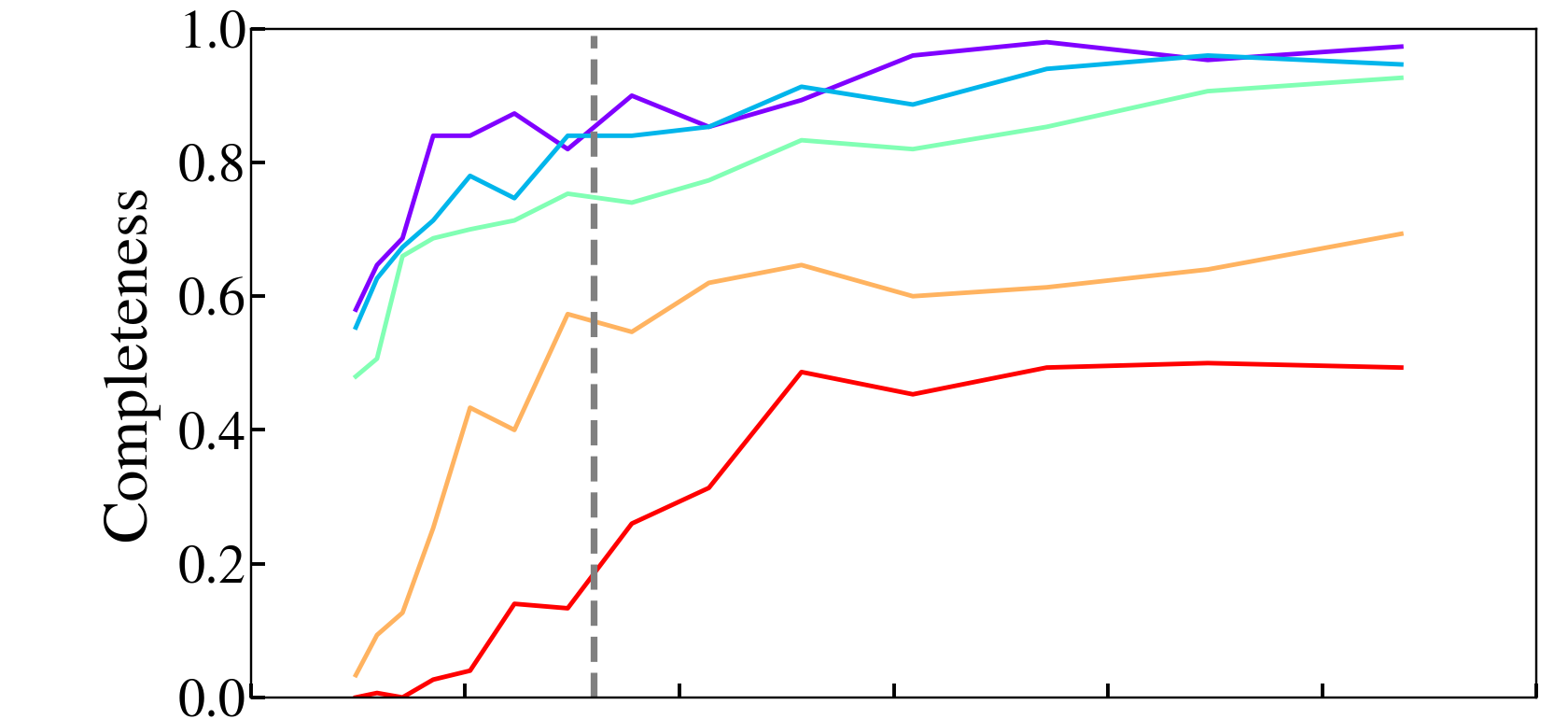}
	\includegraphics[width=0.45\textwidth]{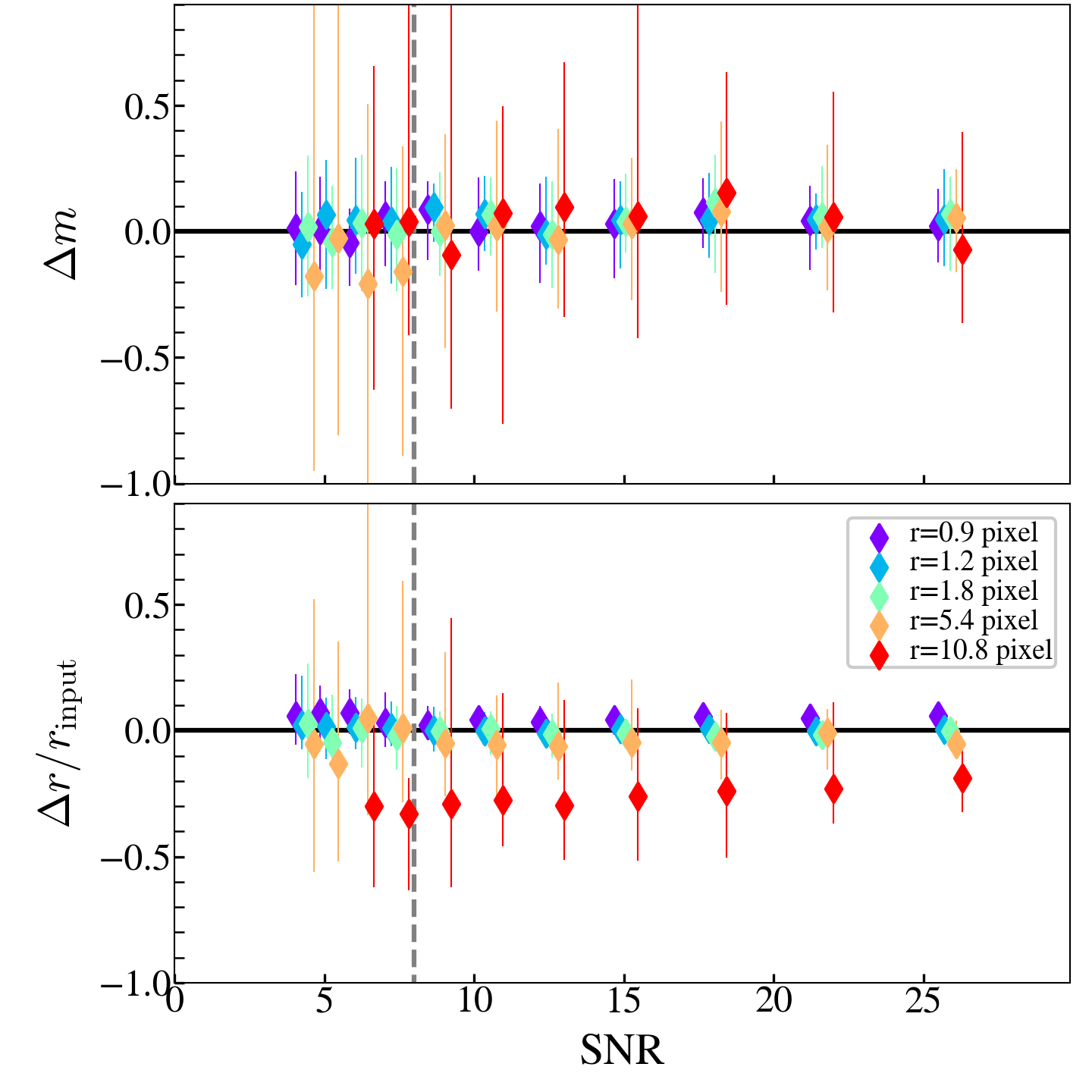}
	\caption{
    {
    (Top): Source completeness as a function of rest-UV SNR, obtained for mock sources of different sizes (solid colored lines). Our conservative SNR cut ($=8$; vertical dashed line) secures high completeness, $>50\,\%$, except for large-sized sources. (Middle): Comparison of magnitudes, $\Delta m = m_{\rm output}-m_{\rm input}$, for sources of different sizes, measured in F150W. Median values (diamonds) and the 16th and 84th percentiles (error bars) are shown.
    (Bottom): Same as the middle panel but for normalized sizes, $\Delta r / r_{\rm input}$.
	}
    }
\label{fig:sizecomp}
\end{figure}

{We investigate the selection completeness of sources with various input sizes and SNRs. We follow a similar procedure presented in \citet{morishita18b} and make mock images by burying artificial sources (of S\'ersic index set to 1) in random, relatively source-free positions in the processed mosaic images. We then repeat our detection and selection analyses as presented in Sec.~\ref{sec:ana} and check their completeness at each SNR and input radius. 

As shown in Fig.~\ref{fig:sizecomp}, the selection completeness is $>50\,\%$ above our SN cut (${\rm SNR_{UV}=8}$) for those with small sizes ($5.4$\,pixel and smaller). The completeness decreases to as low as $\sim20\,\%$ at a larger size. The observed trend is expected, as the source flux is more likely affected by neighbouring sources and affected by background local subtraction and thus cannot be reproduced very accurately. It is also noted that lowering the limiting SNR would not only decrease the completeness but also introduce a lager fraction of low-$z$ interlopers, as reported in \citet[][]{morishita23}.

In the middle and bottom panels of Fig.~\ref{fig:sizecomp}, we show the recovered magnitude and size for the same mock galaxies by our size measurement analysis in the F150W band. 
Above our limiting SNR, magnitudes are overall recovered  well, within $10\,\%$ accuracy. The input size is also recovered  well, down to $\sim0.9$\,pixel, which corresponds to about half of the FWHM size in F150W. This validates our selection criterion used for unresolved sources (Sec.~\ref{sec:compact}). The exception is those with a very large size (10.8\,pixel, or $\sim1.8$\,kpc at $z=7$), where the output size turns to be $30\,\%$ smaller than the input value on average. The trend seen here is overall consistent with the one reported by \citet{ono23}.

While our sample construction has been conducted with an extreme care by maintaining high source completeness, accurate size measurement, and high source purity, the test above implies that our selection may not be as complete at large size, $\simgt2$\,kpc. Improving the completeness in the regime is not impossible but challenging, as detection completeness becomes flattened at a certain SNR and does not linearly improve for large-sized sources. However, we note that $2$\,kpc characterizes $\sim3\,\sigma$ above the median size at $\logm \sim 8.5$, the median stellar mass of our sample galaxies. The distribution of the measured size of our galaxies does not show any evidence of a skewed feature at large radius. As such, if there are any large-sized sources missing in our sample, we speculate that those are still small in number and hardly affect the results of our regression analysis.  
}

\section*{Appendix~D: Image Cutouts of Compact Sources}\label{appen:compact}
Cutout images of the compact sources identified in Sec.~\ref{sec:compact} are shown in Fig.~\ref{fig:stamp}. The only F150W-dropout that is classified as compact is shown in the main text (Fig.~\ref{fig:galfit}).
{We note that completeness for compact sources varies field to field, due to the differences in depth. Indeed, we observe that compact sources are more prevalent in two of the fields, JADESGDS and A2744. This can be explained by depth and magnification, which enhance the identification of low-mass and small-sized sources. We also note that while NGDEEP has a comparable depth as in JADESGDS, it has a relatively small area (single NIRCam pointing). In addition, the NGDEEP field does not allow photometric selections at $z<9.7$, the redshift range where the majority of the compact sources are identified.
}

\begin{figure*}[b!]
\centering
    \input{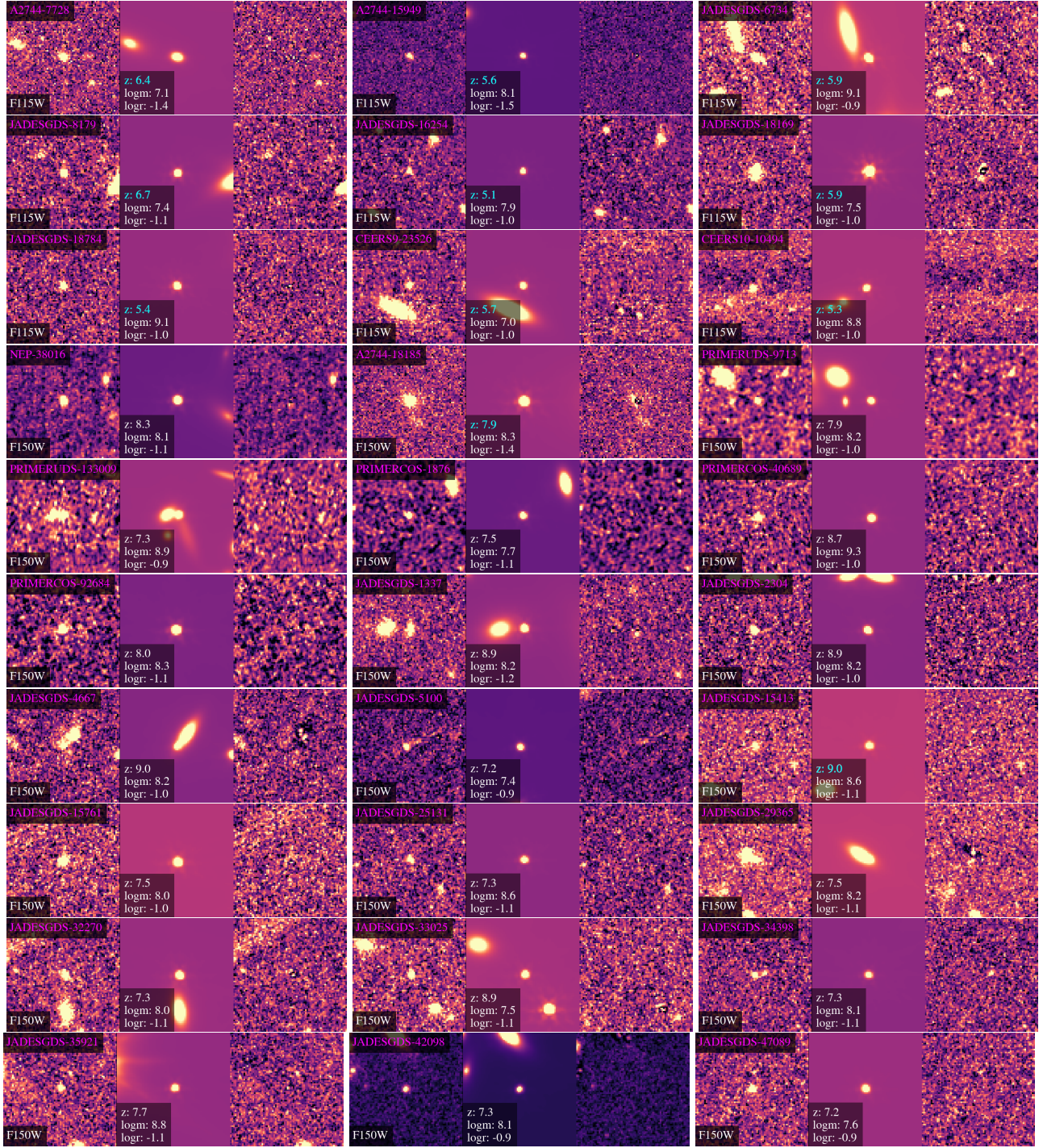}
  	\caption{
    Sources that are classified as compact. ($left$): original image, in the UV band where the size is measured. ($middle$): {\tt galfit} single S\'ersic model image (with S\'ersic index fixed to $1$). ($right$): residual image. Those with spectroscopic redshift measurements are color-coded its redshift label in cyan.
    }
\label{fig:stamp}
\end{figure*}

\begin{figure*}[b!]
\centering
    \input{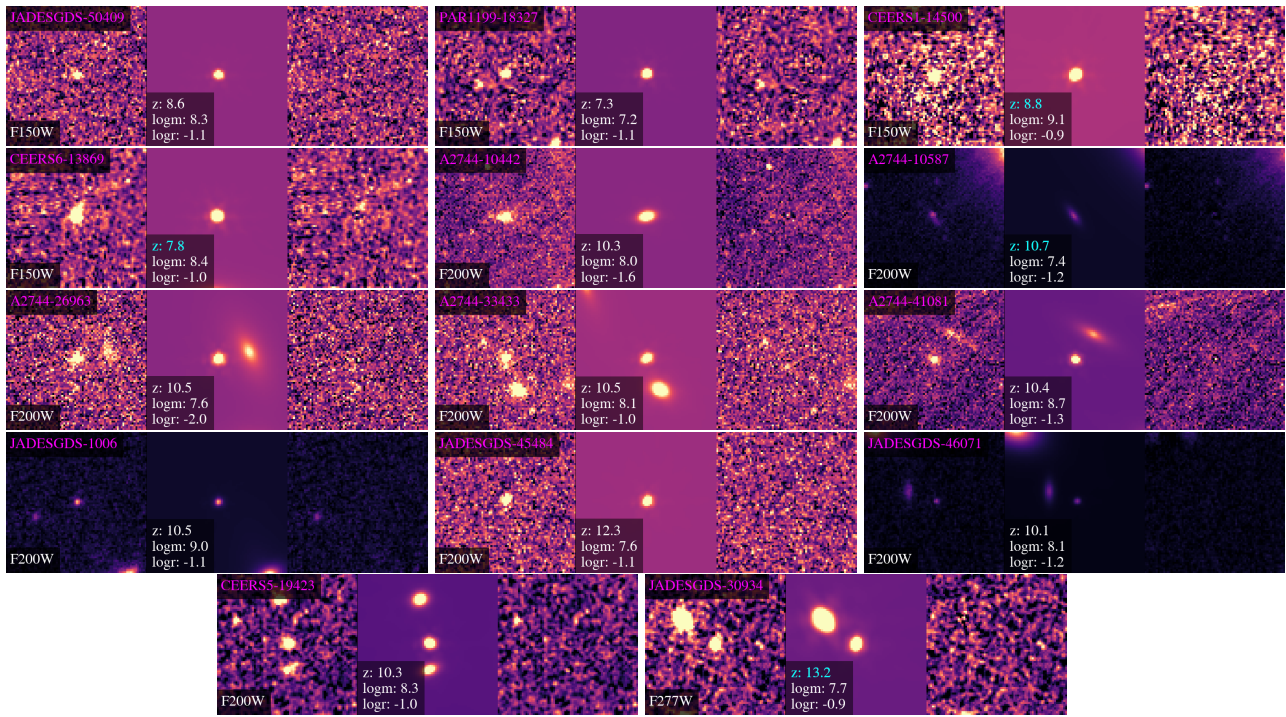}
  	\caption{Continued.
    }
\label{fig:stamp2}
\end{figure*}




\clearpage
\startlongtable
\begin{deluxetable*}{lcccccccccccc}
\tabletypesize{\footnotesize}
\tablecolumns{13}
\tablewidth{0pt} 
\tablecaption{Physical properties of the final candidates.}
\tablehead{
\colhead{ID} & \colhead{R.A.} & \colhead{Decl.} & \colhead{$m_{\rm UV}$} & \colhead{$z^\dagger$} & \colhead{$M_{\rm UV}$} & \colhead{$M_*$} & \colhead{SFR} &\colhead{$\beta_{\rm UV}$} & \colhead{$R_{e,maj}$} &\colhead{\oiii/\hb} & \colhead{$f_{gal}^{\ast}$} & \colhead{$f_{com}^{\ast}$}\\
\colhead{} & \colhead{deg.} &\colhead{deg.} &\colhead{mag} & \colhead{} &\colhead{mag} & \colhead{$\log M_\odot$} & \colhead{$\log M_\odot$/yr} & \colhead{} & \colhead{$\log {\rm kpc}$} & \colhead{$\log$} & \colhead{} & \colhead{}
}
\startdata
\cutinhead{F070W-dropout}
J1235-2632 & 188.92749 & 4.8717985 & $26.7$ & $6.0_{-0.2}^{+0.1}$ & $-19.9_{-0.1}^{+0.1}$ & $8.7_{-0.8}^{+0.8}$ & $0.8_{-0.1}^{+0.1}$ & $-2.7_{-0.1}^{+0.1}$ & $-0.61_{-0.01}^{+0.01}$ & -- & 0 & 0\\
J1235-2875 & 188.94112 & 4.868082 & $27.0$ & $6.4_{-0.2}^{+0.1}$ & $-19.9_{-0.1}^{+0.1}$ & $9.2_{-0.1}^{+0.1}$ & $0.8_{-0.1}^{+0.1}$ & $-2.4_{-0.1}^{+0.2}$ & $-0.65_{-0.01}^{+0.01}$ & -- & 0 & 0\\
J1235-3021 & 188.93146 & 4.8745794 & $27.3$ & $6.0_{-0.2}^{+0.2}$ & $-19.3_{-0.1}^{+0.1}$ & $7.9_{-0.1}^{+0.1}$ & $0.6_{-0.1}^{+0.1}$ & $-2.5_{-0.1}^{+0.1}$ & $-0.56_{-0.02}^{+0.02}$ & -- & 0 & 0\\
J1235-3236 & 188.95332 & 4.8656316 & $27.2$ & $6.8_{-0.2}^{+0.2}$ & $-20.1_{-0.1}^{+0.1}$ & $8.7_{-0.1}^{+0.1}$ & $0.7_{-0.1}^{+0.1}$ & $-2.2_{-0.1}^{+0.1}$ & $-0.90_{-0.02}^{+0.02}$ & -- & 0 & 0\\
J1235-4628 & 188.96568 & 4.8757343 & $25.7$ & $5.2_{-0.1}^{+0.1}$ & $-21.0_{-0.1}^{+0.1}$ & $9.9_{-0.1}^{+0.1}$ & $1.6_{-0.1}^{+0.1}$ & $-1.6_{-0.1}^{+0.1}$ & $-0.27_{-0.02}^{+0.02}$ & -- & 0 & 0\\
J1235-4644 & 188.9286 & 4.8941765 & $27.0$ & $6.3_{-0.2}^{+0.2}$ & $-18.9_{-0.4}^{+0.3}$ & $10.0_{-0.3}^{+0.3}$ & $1.9_{-0.3}^{+0.3}$ & $-0.6_{-0.2}^{+0.5}$ & $-0.46_{-0.03}^{+0.03}$ & -- & 0 & 0\\
J1235-5704 & 188.94789 & 4.8993416 & $26.7$ & $6.6_{-0.1}^{+0.1}$ & $-21.5_{-0.1}^{+0.1}$ & $8.8_{-0.5}^{+0.5}$ & $0.9_{-0.1}^{+0.1}$ & $-2.5_{-0.2}^{+0.2}$ & $-0.84_{-0.01}^{+0.01}$ & -- & 0 & 0\\
J1235-7600 & 188.94092 & 4.915246 & $26.5$ & $5.7_{-0.2}^{+0.2}$ & $-20.1_{-0.1}^{+0.1}$ & $9.4_{-0.1}^{+0.1}$ & $0.9_{-0.1}^{+0.1}$ & $-2.4_{-0.1}^{+0.1}$ & $0.08_{-0.02}^{+0.02}$ & -- & 0 & 0\\
\enddata
\tablecomments{
List of our final sources selected in Sec.~\ref{sec:sample}. Measurements are quoted with $1\,\sigma$ error.
$^\dagger$: Photometric redshift. Errors are omitted for objects with spectroscopic redshift measurements. 
$f_{gal}$: $=1$ when flagged in the {\tt galfit} analysis. $f_{com}$: $=1$ when classified as compact (Sec.~\ref{sec:compact}).
}\label{tab:prop}
\end{deluxetable*}
\clearpage

\bibliography{sample631}{}
\bibliographystyle{aasjournal}



\end{document}